\documentclass[conference]{IEEEtran}

\usepackage[square,sort,compress,comma,numbers]{natbib}

\usepackage{xspace}
\usepackage{filecontents}
\usepackage[draft=true]{hyperref}

\usepackage{balance}

\usepackage{float}
\usepackage{pgf, tikz}

\usepackage{graphicx}

\usepackage[cmex10]{amsmath}

\usepackage{bm}
\usepackage{soul,color} %
\usepackage{amssymb} 

\usepackage{array}

\usepackage{subfigure}
\usepackage{caption2} 

\usepackage{amsfonts}
\usepackage{amsthm}   %

\usepackage{enumerate} % 

\usepackage{extarrows}

\usepackage{tkz-fct}

\newtheorem{theorem}{Theorem} 

\newtheorem{proposition}{Proposition}

\newtheorem{remark}{Remark}

\newtheorem{definition}{Definition}

\newtheorem*{criteria*}{Mode-Selection Criteria}

\usetikzlibrary{patterns,snakes}

\usepackage{enumerate}

\usepackage{extarrows}

\newcommand{\myexpect}[2]{\mathsf{E}_{#1}\left[ #2 \right]}

\newcommand{\snr}{\mathsf{SNR}}
\newcommand{\ssr}{\mathsf{SSR}}

\begin{document}
%\title{\huge Power Allocation for Coding-Free Wireless Networked Control Systems: Water Filling or Fire Quenching?}
\title{ Wireless Networked Control Systems with Coding-Free Data Transmission for Industrial IoT}
%\author{\IEEEauthorblockN{Wanchun Liu, \emph{Member, IEEE}, Petar Popovski, \emph{Fellow, IEEE}, Yonghui Li, \emph{Senior Member, IEEE}, and Branka Vucetic, \emph{Fellow, IEEE}}
%%\thanks{\setlength{\baselineskip}{13pt} \noindent The authors are with School of Electrical and Information Engineering, the University of Sydney, Sydney 2006, Australia.
%%(emails: \{wanchun.liu,\ yc.liang,\ yonghui.li,\ branka.vucetic\}@sydney.edu.au).
%%Part of the paper has been submitted to the Proc. IEEE ICC 2018.
%%}
%}
\author{\IEEEauthorblockN{Wanchun Liu$^\dagger$, Petar Popovski$^\circ$, Yonghui Li$^\dagger$, and Branka Vucetic$^\dagger$}
%	\IEEEauthorblockA{
%	$^\dagger$School of Electrical and Information Engineering, The University of Sydney, Australia\\
%	$^\circ$Department of Electronic Systems, Aalborg University,  Denmark\\
%	Emails:	wanchun.liu@sydney.edu.au,\  petarp@es.aau.dk,\ yonghui.li@sydney.edu.au,\ branka.vucetic@sydney.edu.au
%}
}

\maketitle
\begin{abstract}
\let\thefootnote\relax\footnote{
W. Liu, Y. Li and B. Vucetic are with School of Electrical and Information Engineering, The University of Sydney  (emails: wanchun.liu@sydney.edu.au, yonghui.li@sydney.edu.au and branka.vucetic@sydney.edu.au).
P.~Popovski is with Department of Electronic Systems, Aalborg University (email: petarp@es.aau.dk). Part of the paper has been accepted by Proc. IEEE Globecom 2019~\cite{GC19}. The work of W. Liu and B. Vucetic were supported by Australian Research Council's Australian Laureate
Fellowships Scheme under Project FL160100032.
The work of P. Popovski has been supported the European Research Council (ERC) under the European Union Horizon 2020 research and innovation program (ERC Consolidator Grant Nr. 648382 WILLOW) and Danish Council for Independent Research (Grant Nr. 8022-00284B SEMIOTIC).
The work of Y. Li was supported in part by the Australian Research Council's
Australian Laureate Fellowships Scheme under Project FL160100032 and in part by the Australian Research Council’s Discovery Project Funding Scheme under Project DP150104019. 
}
Wireless networked control systems for the Industrial Internet of Things (IIoT) require low latency communication techniques that are very reliable and resilient.
In this paper, we investigate a coding-free control method to achieve ultra-low latency communications in single-controller-multi-plant networked control systems for both slow and fast fading channels. 
We formulate a power allocation problem 
to optimize the sum cost functions of multiple plants, subject to the plant stabilization condition and the controller's power limit.
Although the optimization problem is a non-convex one, we derive a closed-form solution, which indicates that the optimal power allocation policy for stabilizing the plants with different channel conditions is reminiscent of the channel-inversion policy.
We numerically compare the performance of the proposed coding-free control method and the conventional coding-based control methods in terms of the control performance (i.e., the cost function) of a plant, which shows that the coding-free method is superior in a practical range of SNRs.
\end{abstract}

\begin{IEEEkeywords}
	Low latency communication, optimal power allocation, wireless networked control system, coding-free transmission, analog transmission.
\end{IEEEkeywords}

\section{Introduction}
\subsection{Motivation}

\begin{figure}[t]
	\renewcommand{\captionfont}{\small} \renewcommand{\captionlabelfont}{\small}
	\centering
	%\vspace{-0.3cm}	
	\includegraphics[scale=0.8]{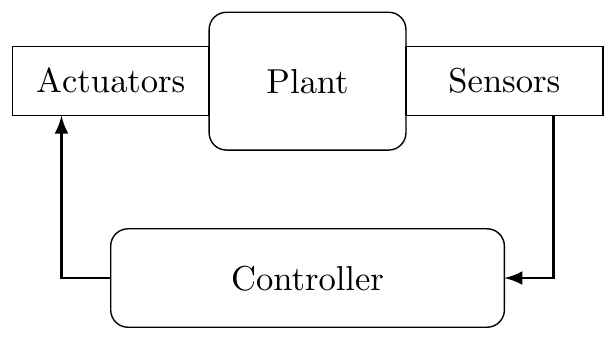}
	\vspace{-0.0cm}
	\caption{A wireless networked control system (WNCS).}
	\label{fig:sys0}
	%\vspace{-0.0cm}
\end{figure}

The industrial Internet of Things (IIoT) is the concept of using smart sensors and actuators to enhance industrial and manufacturing processes~\cite{Galloway}. 
Different from consumer IoT applications, the IIoT is more focused on industrial control ranging from building and process automation to the more critical scenarios in power systems automation and power electronics control~\cite{Pang}.
Wireless networked control systems (WNCSs) have been considered as a key technical solution for flexible deployment of industrial control, and have attracted a lot of attention from both industry and academia~\cite{ParkSurvey}.
%
%Driven by recent significant progress in wireless communications, networking, sensing, computing,
%and control, as well as their applications in the Industrial Internet-of-Things (IIoT)~\cite{IIOT}, wireless networked control systems (WNCSs) have attracted a lot of attention from
%both industry and academia~\cite{ParkSurvey}. 
In general, a WNCS consists of a dynamic unstable plant (e.g., a chemical/power plant, robot or unmanned aerial vehicle) to be controlled, sensor nodes that measure and report the plant state, a remote controller that receives the sensors' measurements and sends control signals, and actuators that receive the control signals to control the plant, as illustrated in Fig.~\ref{fig:sys0}.

Different elements in WNCSs communicate  with each other through wireless networks. 
Most of the existing work in WNCSs only considered coding-based digital communication systems for sensor-controller and controller-actuator communications~\cite{ParkSurvey}. 
Although digital communications have been widely adopted in cellular networks and  wireless local area networks for high-speed large-volume digital data transmissions, extensive research has focused on coding-free analog communications in the areas of wireless sensor networks (WSNs) including wireless remote estimation systems, which have closer ties with the analog physical world (see \cite{Jiang1,Jiang2,Alex,Cui,AlexW} and the references therein).
In particular, for the special case where both the source signal and the channel are Gaussian, it has been proved that coding-free analog signaling is the optimal in terms of distortion between the source signal and the recovered signal at the receiver~\cite{Gastpar,Gastpar2,Gastpar3}.
%
%This is mainly because it has been proved that coding-based communications, which require signal quantization and digital source and channel encoding, achieve an exponentially worse performance than coding-free analog signaling in terms of distortion between the source signal and the recovered signal at the receiver~\cite{Gastpar,Gastpar2,Gastpar3}.
Although coding-free communications have been extensively studied in WSNs, they have rarely been investigated for WNCSs in the open literature.
Note that WSN and WNCS work in very different ways. The former only needs to estimate the source signal through wireless channels, i.e., an open-loop system, while the latter further requires to generate and send control commands to actuators through wireless channels to control the source signal, i.e., a closed-loop system. Thus, the design of coding-free communications can be very different in WSN and WNCS.
In this work, we ask a fundamental question: 
\emph{Which communication method does lead to a better control performance of WNCS, coding-based or coding-free one?}

%The symbol-level analog communication scheme has ultra-low latency and complexity compared to conventional coding-based communications that send coded commands with many channel uses and adopt complex decoding algorithms (e.g., maximum likelihood detection and decoding). 
%Certainly, coding-free communications cannot guarantee noiseless reception of the original information, as there always exists a fundamental tradeoff between transmission latency and reliability in communication systems~\cite{Petar,Mahyar}.
%For a WNCS, only the long-term control performance of the plant is of interest~\cite{ParkSurvey,Nair2004,TCP}, 
%thus, it is important to investigate whether the one using coding-free transmission, which provides short transmission latency but noisy control signals, can beat the one using coding-based transmission, which introduces longer latency but provides more accurate control commands, in terms of the control performance of WNCSs.

%Different from conventional NCSs that adopt wired communications,
%a WNCS transmits sensors' measurement and the control signals through wireless channels, which generally offer lower SNR and are highly variable.

\subsection{Related Work}
\emph{\underline{Coding-based communications in WNCSs}}. In the literature on WNCSs, there are various types of wireless digital communication system modeling for the sensor-controller and controller-actuator channels.
Assuming a communication system that is noiseless and has a limited data rate, the minimum required data rate to stabilize the plant was obtained in~\cite{Mitter2004,Nair2004,Savkin2004}.
Assuming arbitrarily large number of quantization levels and high coding rate, but independent and identically distributed (i.i.d.) packet dropouts, the optimal control law and the stability condition of WNCS in terms of the packet dropout probability were derived in~\cite{TCP} and the references therein. Based on such pioneering works, the results have been extended to
the settings of multi-sensor scheduling in WNCSs~\cite{Alex2,Shi}, and
multi-WNCS scheduling over shared wireless communication resources~\cite{Gatsis}.
More recently, the packet-dropout model has been adopted in WNCSs with security constraints~\cite{secure1,secure2,secure3,secure4}.
Considering both random packet delays and packet dropouts, an optimal state estimation problem of WNCS was investigated in~\cite{Randomdelay}.

\emph{\underline{Coding-free communications}}.
In general, coding-free transmission in WSNs means that each sensor merely
transmits a signal proportional to its observations, without
any quantization and further coding~\cite{Gastpar}.
Thus, coding-free scheme has ultra-low latency and complexity compared to conventional coding-based communications that send coded information with many channel uses and adopt complex decoding algorithms (e.g., maximum likelihood detection and decoding). 
Inspired by the pioneering work~\cite{Gastpar}, a number of related research works using coding-free transmissions emerged in recent years~\cite{Jiang1,Jiang2,Alex,Cui,AlexW,AirComp,consensus,GuangxuWangyong,Gunduz,Osvaldo}.
In \cite{Jiang1,Jiang2,AlexW,Cui}, the problems of deterministic parameter estimation through multiple-access channels were comprehensively investigated.
In~\cite{Alex}, the optimal power allocation problem of a remote estimation system with a Gaussian-Markov source was investigated.
More recently, coding-free communications were adopted for effective data fusion of a massive number of sensors in high-mobility sensing applications in~\cite{AirComp}.
Coding-free communications were also adopted in multi-agent systems for achieving global consensus of the agents status~\cite{consensus}.
In~\cite{GuangxuWangyong,Gunduz,Osvaldo}, coding-free communication schemes were applied in cellular systems for fast parameter exchanging in online training processes of wireless machine-learning applications.

%It is worth mentioning that coding-free communications itself is not new and has been considered for different applications in addition to WNCSs. 

\subsection{Contributions}

%...Note that all these models of wireless digital communication
%systems adopted in WNCSs are only valid under restrictive assumptions and not unrealistic.
%In practice, a digital communication system performs quantization, source coding and channel coding for transmission.
%Such a communication system has three performance parameters, i.e., latency, data rate and reliability. Each parameter has direct effect on the performance of WNCS.
%However, 
%
%
%...assumptions and do not simultaneously consider the effect of
%the latency, the data rate and the reliability, i.e., the three
%performance parameters of a high fidelity model for practical
%wireless communication systems [10].
%
%......Most of these works (implicitly) rely on joint source-channel coding methods for data transmissions in WNCSs. However, such coded data transmission methods introduce latency in delay-sensitive WNCSs due to the source-channel coding blocklength. Furthermore, the performance of a WNCS with practical coded transmissions including quantization and channel coding remains largely unknown, making the optimal design of WNCS very challenging~\cite{ParkSurvey}.

In this work, we investigate coding-free WNCSs inspired by the existing works on coding-free communications~\cite{Jiang1,Jiang2,Alex,Cui,AlexW,Gastpar,Gastpar2,Gastpar3,AirComp,consensus,GuangxuWangyong,Gunduz,Osvaldo}.
To be specific, we focus on the wireless controller-actuator channel, where the control signal generated by the \emph{controller} is an amplitude-modulated-like analog signal, and the \emph{actuator} linearly scales the received control signal and applies it to the plant directly.\footnote{Although coding-free communication method itself is not new, integrating coding-free communications with WNCSs introduces new design problems with different optimizing targets and constraints from the ones considered in coding-free communication systems~\cite{Jiang1,Jiang2,Alex,Cui,AlexW,Gastpar,Gastpar2,Gastpar3,AirComp,consensus,GuangxuWangyong,Gunduz,Osvaldo}.}

%Certainly, analog communications cannot guarantee noiseless reception of the control data, as there always exists a fundamental tradeoff between transmission latency and reliability in communication systems~\cite{Petar,Mahyar}.
%For a WNCS, only the long-term control performance of the plant is of interest~\cite{ParkSurvey,Nair2004,TCP}, 
%thus, it is important to investigate whether the one using coding-free transmission, which provides short transmission latency but noisy control signals, can beat the one using coding-based transmission, which introduces longer latency but provides more accurate control commands, in terms of the control performance of WNCSs.
%	However, by integrating it with emerging telecommunication techniques, such as massive MIMO,
%it can also benefit from multiple paths and channel hardening~\cite{massive}.

Note that due to the fundamental tradeoff between transmission latency and reliability in communication systems~\cite{Petar,Mahyar}, such the symbol-level coding-free communications significantly reduce the latency 
but cannot guarantee noiseless reception of the control signal, while coding-based communications can achieve a much reliable control with longer latency.
For a WNCS, only the long-term control performance of the plant is of interest~\cite{ParkSurvey,Nair2004,TCP}, which is a distortion-type measurement, and coding-free transmission may beat coding-based transmission in terms of the control performance of WNCSs.
%Coding-free transmission, which provides short transmission latency but noisy control signals, may beat coding-based transmission, which introduces longer latency but provides more accurate control commands, in terms of the control performance of WNCSs.

As presented earlier the models of coding-based communication
systems adopted in most of the WNCS works are only valid under restrictive assumptions and not unrealistic~\cite{Mitter2004,Nair2004,Savkin2004,TCP,Alex2,Shi,Gatsis,secure1,secure2,secure3,secure4,Randomdelay}.
In practice, a coding-based communication system performs quantization, source coding and channel coding for transmission.
To the best of our knowledge, the control performance of a coding-based WNCS with practical quantization, source and channel coding through noisy wireless channel remains largely unknown, making the optimal design of coding-based WNCS very challenging~\cite{ParkSurvey}.
In this work, we provide a comprehensive performance analysis and optimization for coding-free WNCSs, and numerically compare the control performance of coding-based and coding-free WNCSs.

The main contributions of the paper are summarized as follows:
%\textbf{Contributions of this paper}: 
\begin{itemize}
	\item We consider a WNCS consisting of a single controller and a single or multiple plants, and propose a joint digital-analog wireless control protocol, where a digital header is preserved to support networking functions, and the data transmission for plant control is analog.
	\item For the coding-free control process of the single plant scenario, we propose to jointly optimize the parameters of the controller and the actuator to minimize the average cost function of the dynamic plant subject to the transmission-power constraint of the controller and the stability condition of the plant. This is done for both slow and fast fading channels.
	Although the original problem is non-convex, we derive the optimal parameters in a closed form. To the best of our knowledge, this problem has never been considered in the literature. 
	Furthermore, we have derived the theoretical condition on the existence of a coding-free control protocol that can stabilize the plant, in terms of the transmission-power limit and the channel conditions.
	\item We extend the single-plant results to the multi-plant scenario. Specifically, we formulate a 
	novel optimal transmission power allocation problem among multiple plants under the total transmission power constraint of the controller such that each plant is stabilized and the sum cost functions of the plants is minimized as well. Our results show that the optimal power allocation policy is of a channel-inversion type, i.e., we need to allocate more power for the worse channel conditions.
	\item We numerically compare the performance of practical coding-based WNCS with coding-free WNCS. Our results shows that the coding-free method is superior in a practical range of SNRs.
\end{itemize}

%We also would like to clarity the main differences between our work and a recent work considering coding-free control, i.e., \cite{Lau2}.
%First, the system and network models are different. In~\cite{Lau2}, a coding-free method has been considered for the transmissions between sensors and a controller, i.e., the sensing-feedback channel,
%and a novel multi-access protocol was proposed for data transmissions.
%However, we focus on the coding-free transmission between the controller and multiple actuators, i.e., the control channel, and consider a broadcasting network that the single controller transmits control signal to different actuators simultaneously.
%Second, the performance metrics/analysis are different. In~\cite{Lau2}, the stability analysis of the plant and the mean square error of state estimation at the controller were delivered, while we analyze the average cost function of the plant and then minimize it.
%Last, we consider a hybrid digital/analog encoding, where the digital one is used to support networking functions. This makes the protocol closer to practice, which was not considered in~\cite{Lau2}.

The remainder of this paper is organized as follows.
Section~\ref{section2} presents the proposed WNCS with coding-free data transmission.
Sections~\ref{sec:single} and \ref{sec:multi} investigate the optimal coding-free control methods for single- and multi-plant networks, respectively, under slow-fading channel. 
Section~\ref{sec:fast} investigates the optimal coding-free control methods for single- and multi-plant networks, under fast-fading channel. 
Section~\ref{sec:num} numerically presents the performance of the proposed coding-free control method. 
Finally, Section~\ref{sec:con} concludes the paper.

%Different from~\cite{SNR} and~\cite{Lau}, which focused on the sensor-to-controller channels and single-controller-single-plant scenarios, we consider the design of the communication on the controller-to-actuator channel and formulate a 
%novel optimal transmit power allocation problem among multiple plants under the total transmit power constraint of the controller such that each plant is stabilized and the sum cost functions of the plants is minimized as well.
%It is interesting to see that the optimal power allocation policy is of a channel-inversion type, i.e., we need to allocate more power for sending the control signal to the plant with a worse channel condition.

%Assuming a delay-free Gaussian continuous channel, i.e., the wireless communication system has infinite number of channel inputs, the stability condition in terms of SNR has been investigated in~\cite{Tamer} and \cite{SNR}.
%

\begin{figure}[t]
	\renewcommand{\captionfont}{\small} \renewcommand{\captionlabelfont}{\small}
	\centering
	%\vspace{-0.3cm}	
	\includegraphics[scale=0.7]{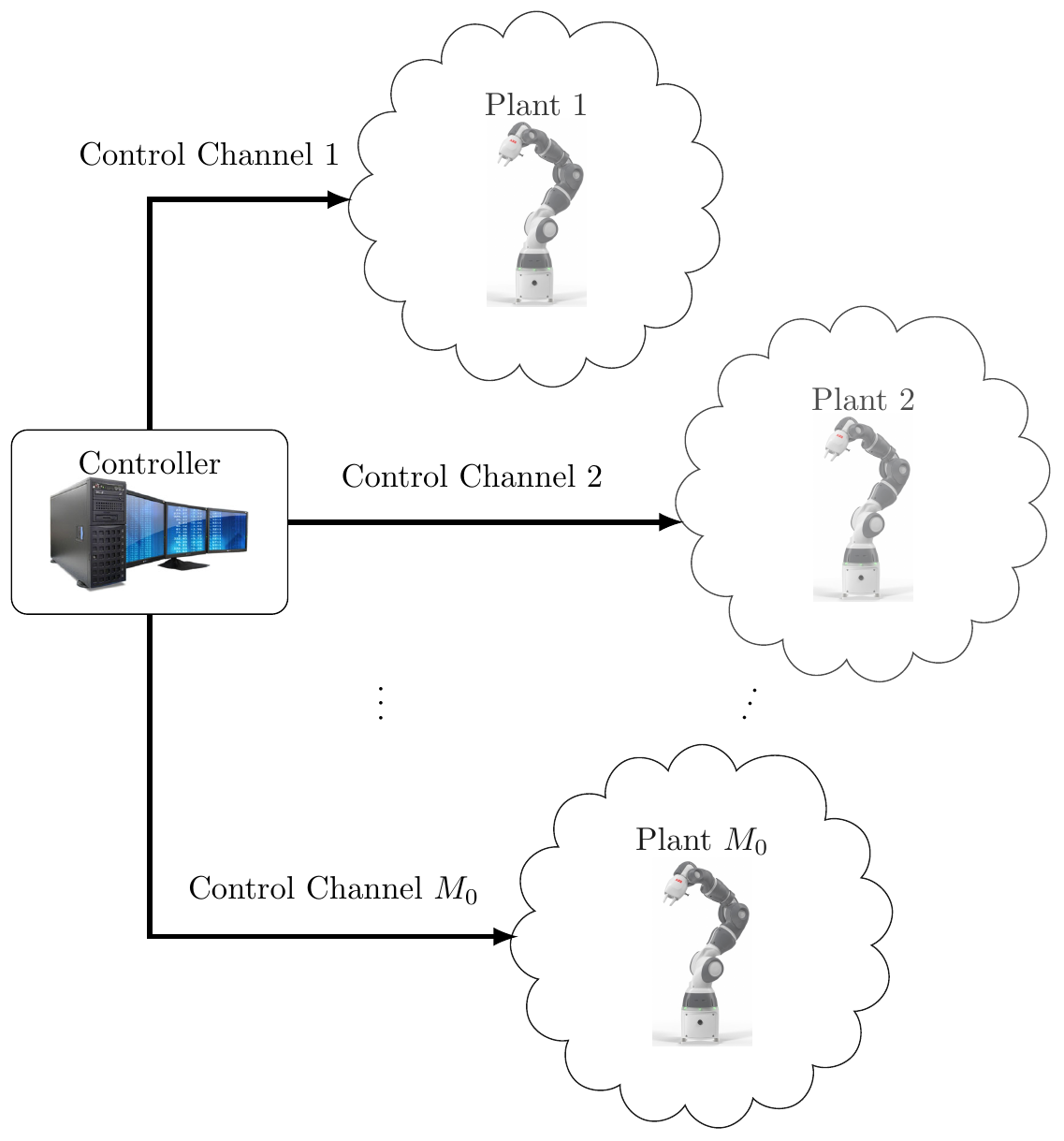}
	\vspace{-0.0cm}
	\caption{Illustration of an $M_0$-plant WNCS.}
	\label{fig:sys}
	%\vspace{-0.0cm}
\end{figure}
\section{System Model}\label{section2}
We consider a WNCS with a single controller and $M_0$ remote plants each equipped with an actuator to be controlled, as illustrated in Fig.~\ref{fig:sys}.
It is assumed that the plant states are perfectly known by the controller, i.e., the sensors that measures the plant states are co-located with the controller~\cite{dead,QuevedoGaussian}.\footnote{Note that there are many applications where the sensors that measure the plants' state and the controller are collocated. For example, the controller observes the movement and location of automated guided vehicles (AGVs) and robots by using its cameras (treated as sensors), see e.g.,~\cite{Sergey}, and send control commands to the AGV or robots through wireless channels.}
Based on the current plant states, the controller is able to generate and send control signals to the $M_0$ actuators simultaneously through $M_0$ independent channels to control the plants~\cite{Gatsis}.
%The controller sends control signals to each actuator on the plant based on the plant states measured by the sensors.
We assume that each plant has one real state (i.e., the forward velocity of a mobile robot or the force of a robot gripper) to be controlled, i.e., a scaler system~\cite{Sahai,Nair09,Serdar}.\footnote{A more general vector system is beyond the scope of the paper.}

%We consider a slowly changing environment.
%The effective channel coefficient of the link between the controller and actuator $i$ is denoted by $H_i \in \mathbb{C}, \forall i$. 

\subsection{Coding-Free Control Protocol}
\begin{figure}[t]
	\renewcommand{\captionfont}{\small} \renewcommand{\captionlabelfont}{\small}
	\centering
	\includegraphics[scale=1.0]{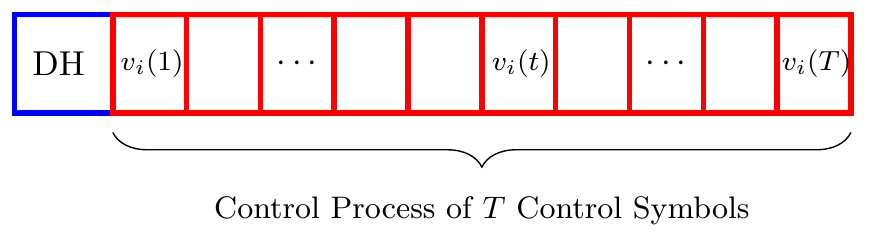}
%	\vspace{-0.0cm}
	\caption{Coding-free control protocol for plant $i$.}
	\label{fig:protocal}
%	\vspace{-0.5cm}
\end{figure}

In general, the proposed coding-free control protocol consists of two phases: digital header (DH) transmission and coding-free control process, as illustrated in Fig.~\ref{fig:protocal}.
Before starting each control process (i.e., sending control signals to the actuators so as to control the plants), the controller needs to send a DH to each plant to perform networking functions, such as identification and authentication to avoid the situation in which the plants received commands from malicious transmitters.

%The $M$ plants that have successfully detected the header, can authenticate the controller and thus be selected for coding-free control, as they are less likely to suffer from environmental interferences.
%The remaining $(M_0-M)$ actuators can switch to a predetermined self-control mode, e.g., implementing zero input, or repeat the previous control action, or predict the control signal to be implemented based on the previous ones (see~\cite{ParkSurvey} and the references therein), and wait for the next transmission round.
%The reliability of the header transmission can be found in~\cite{Petar}.

%Moreover, by adopting massive MIMO, a diversity technique, the channel can be treated as a Gaussian channel due to the channel hardening phenomenon~\cite{massive}. 
%\emph{The fading channel scenario can be considered in our further work, where a large sequence of coherence time are need to stabilize the plant.}

Since the channel between the controller and an actuator can be in deep fading, only a set of plants with good channel conditions are chosen by the controller for \emph{remote coding-free control}. The other plants can switch to a predetermined \emph{self-control mode}, e.g., repeating the previous control action, or predicting the control signal to be implemented based on the previous ones (see~\cite{ParkSurvey,phdthesis} and the references therein), and wait for the next control process.

{By using the property of channel reciprocity and the classical channel estimation method~\cite{BOOKTse}, i.e., letting the actuators sending pre-known pilot symbols to the controller, the controller is able to acquire the qualities of the $M_0$ controller-actuator channels,
based on which it determines the set of plants for coding-free control.
Note that the number of plants selected, $M\leq M_0$, may change in different control processes depending on the corresponding channel conditions. 
The selection criteria of the plants will be given in Sections~\ref{sec:single}-\ref{sec:multi} and \ref{sec:fast} for single- and multi-plant cases under slow and fast fading channels, respectively. 
Note that the control-mode information of each plant is also contained in the DH to inform the plants before a control process, and the DH contains both the original information bits and their cyclic redundancy check (CRC) for error detection~\cite{CRC}.
Once an actuator receives a DH, it calculates the CRC based on the received information bits. If the calculated CRC matched the received CRC, the actuator sends a one-bit acknowledgment (ACK) signal to the controller. Otherwise, a detection error occurs and it sends a negative-acknowledgment signal alternatively.
We assume that a DH can be detected correctly by the plants selected for coding-free control, which have relatively good channel conditions, since the DH only contains a few information bits. Once a detection error occurs, we can do retransmission for the DH, which does not introduce much delay as the DH is short.
If a plant that is not selected for coding-free control, cannot detect the DH correctly, it can autonomously switch to the self-control mode.}
\emph{In this work, we only focus on the plants selected for coding-free control.}

A control process contains $T$ control symbols as illustrated in Fig.~\ref{fig:protocal}.
At discrete time $t \leq T$, where $t$ can be treated as the symbol index, each plant state, $x_i(t)\in \mathbb{R}, \forall i \in \{1,\cdots,M\}$, can be obtained by the controller~\cite{dead,QuevedoGaussian}, {which then performs a multiple unicast over $M$ independent channels to send the control signals to $M$ dedicated actuators. Note that information transmission through multiple parallel channels has been systematically investigated in~\cite{Broadcast} and the references therein. }
There is a common power constraint, $\mathsf{P}_0$, for all transmission channels.
The controller adopts a \emph{linear control policy}~\cite{dead}, and the control signal sent by the controller for plant~$i$ at time~$t$ is given~by
\begin{equation} \label{v}
v_i(t) = K_i x_i(t),
\end{equation}
where the controller factor, $K_i \in \mathbb{R}$, is a design parameter.

\subsection{Wireless Communication Channels}
We consider two types of channel models: {slow-fading channel} and {fast-fading channel}.

For the \emph{slow-fading channel} case, the channel coefficient of the link between the controller and actuator $i$, $H_i\in \mathbb{C}$, is fixed during a DH and a control process due to the assumed long coherence time. 
By using the classical channel estimation method~\cite{BOOKTse}, we assume that $H_i$ is available at the controller and actuator $i$, $\forall i  \in \{1,\cdots,M_0\}$ \cite{Gatsis,Gatsis2}.
In this case, the controller controls the plants through parallel Gaussian channels and aims for driving the plant state close to the steady state in the control process of $T$ symbols. This channel model is suitable for low-mobility industrial control applications. For instance, the symbol time is $4~\mu$s of the IEEE 802.11 standard. Since the typical Doppler shift of a industrial automation factory is about $10$~Hz~\cite{thesis}, the
typical channel coherence time is about $100$~ms. As each symbol carries one control signal, we have tens of thousands of control actions within one channel coherence time, which will be illustrated in Sec.~\ref{sec:num_process}.
%Furthermore, due to the phenomenon of channel hardening, when the number of the controller's antennas is large (e.g., $>200$), the fading channel between the controller and a actuator approaches to a static channel~\cite{massive}.

For the \emph{fast-fading channel} case, the channel coefficient of the link between the controller and actuator $i$, $H_i(t)\in \mathbb{C}$, changes symbol by symbol and follows a Gaussian distribution. 
We assume that the variances of the $M_0$ channels, $\sigma^2_{h,i},\forall i \in \{1,\cdots,M_0\}$, are known by the controller and the actuators.
Similar to the slow-fading case, the controller aims for driving the plant state close to the steady state in the control process of $T$ symbols with different channel conditions.
This channel model is suitable for high-mobility industrial control applications.

\subsection{Control Action}
{Given the control signal sent by the controller, $v_i(t)$, the received signal at actuator $i$ through the fading channel is given~as
\begin{equation}
\tilde{r}_i(t)=\begin{cases}
 H_i v_i(t) + \tilde{z}_i(t) ,&\text{slow fading}\\
 H_i(t) v_i(t) + \tilde{z}_i(t) ,&\text{fast fading}
\end{cases}
\end{equation}
where $\tilde{z}_i(t)$ is the complex additive white Gaussian noise (AWGN).
Since the transmitted signal $v_i(t)$ is real, actually only one dimension of the complex channel is utilized. Specifically, for the slow fading scenario, since $H_i \triangleq \vert H_i \vert e^{j\phi_i}$ is known to actuator $i$, the received control signal can be rewritten as
\begin{equation}
\tilde{r}_i(t) = \vert H_i\vert e^{j\phi_i}  v_i(t) + z^{\phi_i}_i(t) e^{j\phi_i} + z^{\phi_i\perp}_i(t) e^{j(\phi_i+\pi/2)},
\end{equation}
where $z^{\phi_i}_i(t)$ and $z^{\phi_i\perp}_i(t)$ are real signals obtained by projecting $\tilde{z}_i(t)$ on the orthogonal bases of $e^{j\phi_i}$ and $e^{j(\phi_i+\pi/2)}$, respectively. Since the original control signal is projected on the base of $e^{j\phi_i}$, we only need to consider the received real signal on the same base, i.e.,
\begin{equation}\label{slow}
r_i(t)= \vert H_i \vert v_i(t)+z^{\phi_i}_i(t).
\end{equation}
For the fast fading scenario with $H_i(t)\triangleq \vert H_i(t)\vert e^{j\phi_i(t)}$, since the phase information $\phi_i(t)$ changes with time and is unknown to actuator $i$, it is not possible to determine the base of $e^{j\phi_i(t)}$ carrying the real signal $v_i(t)$. For simplicity, we assume that $\phi_i(t)=0$ and hence only consider the real part of the received signal, i.e.,
\begin{equation}\label{fast}
r_i(t) = \text{Real}[\tilde{r}_i(t)] = \text{Real}[H_i(t)] v_i(t) + \text{Real}[\tilde{z}_i(t)].
\end{equation}
From \eqref{slow} and \eqref{fast}, the effective (real) received control signal in the slow and fast fading scenarios can be unified as
\begin{equation}
r_i(t)=\begin{cases}
 H_i v_i(t) + z_i(t) ,&\text{slow fading}\\
H_i(t) v_i(t) + z_i(t) ,&\text{fast fading}
\end{cases}
\end{equation}
where $H_i\in \mathbb{R}^+$ and $H_i(t)\in \mathbb{R}$, and the noise $z_i(t)$ is a real AWGN with a zero mean and variance~$\sigma^2_z$.

Then, the implemented control signal by actuator~$i$ is simply a linear mapping of the received control signal as
\begin{equation} \label{u}
u_i(t) =  G_i  r_i(t),
%\begin{cases}
%G_i \left( H_i v_i(t) + z_i(t) \right),&\text{slow fading}\\
%G_i \left( H_i(t) v_i(t) + z_i(t) \right),&\text{fast fading}
%\end{cases}
\end{equation}
where the actuator factor, $G_i\in \mathbb{R}$, is a design parameter.}

\subsection{Plant Under Control and Performance Metrics} \label{sec:metric}
The linear discrete dynamic process of the $i$th plant's state is presented as (see e.g., \cite{dead})
\begin{equation} \label{P}
\begin{aligned}
x_i(t+1) &= A x_i(t) + u_i(t) + w_i(t),
\end{aligned}
\end{equation}
where $w_i(t)$ is the plant disturbance, which is an i.i.d. real Gaussian process with zero mean and variance $\sigma^2_w$.
The real number $A$ is the plant parameter and it is the same for all plants, though the results obtained in the rest of the paper can be readily
extended to settings with different plant parameters.
From~\eqref{P}, we see that the next plant state depends on the current state, implemented control action and plant disturbance.
Also, to avoid a trivial problem, we assume that the plant is \emph{open-loop unstable} by letting $\vert A \vert >1$, i.e., the plant state grows unbounded in the absence of control inputs, i.e., $u_i(t)=0,\forall t$.

{The sum cost of the $M$ plants for the $T$ time slots of the control process is defined in a quadratic form as
\begin{equation}\label{def:cost}
J_T \triangleq \frac{1}{T} \sum_{t=1}^{T} \sum_{i=1}^{M} \myexpect{\!}{\left(x_i(t)\right)^2}.
\end{equation}
Note that such a type of quadratic cost is most commonly used in the literature of WNCSs~\cite{dead,TCP,Shi,Alex2,Gatsis,Gatsis2,Randomdelay,Nair2004,Nair09}, which is mainly because quadratic function is convex and smooth, and facilitates the analysis and optimization of the performance of WNCS.  }

The mean-square stability condition is
\begin{equation}\label{def:cost_longterm}
J_{\text{ave}} \triangleq \limsup\limits_{T\rightarrow \infty} J_T < \infty.
\end{equation}
In other words, the $M$-plant system is stabilized by the remote controller if the long-term average cost $J_{\text{ave}}$ is bounded.

In practice, a control process always has a certain deadline $T$, and the cost $J_T$ is always bounded in this sense. However, the long-term stability condition~\eqref{def:cost_longterm} and the long-term cost $J_{\text{ave}}$ are useful and commonly adopted in the literature see e.g.,~\cite{Nair2004,TCP}.
This is because in the large $T$ scenario (i.e., in the order of a few hundred), if the condition~\eqref{def:cost_longterm} is satisfied, $J_T$ is quite close to the long-term performance $J_{\text{ave}}$; otherwise, $J_T$ is huge as $\sum_{i=1}^{M}\myexpect{\!}{\left(x_i(t)\right)^2}$ grows exponentially with time in the unstable scenario, which will be illustrated in Sec.~\ref{sec:num_process}.

\emph{Therefore, in what follows, we focus on the optimal design of the controller and actuator factors, $K_i$ and $G_i$, such that the long-term cost of the plants, $J_{\text{ave}}$, is bounded and minimized under the total power constraint of the controller.}
%Note that $J_{\text{ave}}$ is a good approximation of $J_T$ when the number of control symbols in a control process, $T$, is large and within a practical range, which will be verified in Sec.~\ref{sec:num}.

\section{Slow-Fading-Single-Plant Case: Analysis and Optimization}\label{sec:single}
{In this section, we consider the optimal design problem of the single-plant case in slow-fading channels, and hence drop out the plant index $i$ of $x_i(t)$, $w_i(t)$, $z_i(t)$, $v_i(t)$, $H_i$, $K_i$ and $G_i$ in the following analysis.}

Taking \eqref{v} and \eqref{u} into \eqref{P}, the dynamic process of the plant state can be rewritten as
\begin{equation} \label{x-x}
\begin{aligned}
x(t+1) 
&= A_c x(t) +  G z(t) + w(t),
\end{aligned}
\end{equation}
where
\begin{equation}
A_c \triangleq A +  G H K
\end{equation}
is the closed-loop plant parameter. 

Taking expectation on both sides of \eqref{x-x}, we have
\begin{equation} \label{Ex}
\myexpect{\!}{x^2(t+1)}=A^2_c \myexpect{\!}{x^2(t)} + G^2 \sigma^2_z + \sigma^2_w.
\end{equation}
To make the long-term average cost $J_\text{ave}$ in \eqref{def:cost_longterm} bounded, $\myexpect{\!}{x^2(t)}$ should be bounded when $t\rightarrow \infty$.
From \eqref{Ex}, it is clear that $\myexpect{\!}{x^2(t)}$ is bounded iff $A^2_c<1$. Thus, the stability condition of the plant in the sense of \eqref{def:cost_longterm} is obtained as below.

\begin{definition}[Stability condition in slow-fading channel]
	\normalfont
The plant \eqref{P} is closed-loop  mean-squared stable using the coding-free control method \eqref{v} and \eqref{u} iff 
\begin{equation} \label{Ac}
\vert A_c \vert < 1.
\end{equation}
\end{definition}
Therefore, to stabilize the plant, \emph{it is assumed that $A > 0$, $H >0$, $G >0$ and $K <0$ for brevity in the rest of the paper without loss of generality.}

%Using the updating rule of the state variance \eqref{Ex}, we further have
%\begin{equation} \label{x-01}
%\myexpect{\!}{x^2(t)}= \left({A^2_c}\right)^t \myexpect{\!}{x^2(0)} + \sum_{j=0}^{t-1} \left({A^2_c}\right)^{t-1-j} \left(G^2 \sigma^2_z + \sigma^2_w\right).
%\end{equation}

%
From \eqref{x-x}, it is straightforward that
\begin{equation} \label{x-0}
x(t) = A^t_c x(0) + \sum_{j=0}^{t-1} A^{t-1-j}_c \left(G z(j) + w(j)\right).
\end{equation}
As we consider the long-term performance of the closed-loop stable dynamic process, 
letting $t \rightarrow \infty$, we have ${A^t_c} {x(0)} \rightarrow 0$ in \eqref{x-0}, and the steady-state distribution of $x(t)$ is zero-mean Gaussian with
the variance
\begin{equation} \label{average_x2}
\myexpect{\!}{x^2(t)} = \frac{G^2 \sigma^2_z + \sigma^2_w}{(1-A^2_c)},
\end{equation}
and hence
\begin{equation}\label{varience_x}
J_{\text{ave}} = \myexpect{\!}{x^2(t)}.
\end{equation}
From \eqref{v} and \eqref{average_x2}, the steady-state distribution of the controller signal, $v(t)$, is also zero-mean Gaussian with the variance
\begin{equation}
\mathsf{P} \triangleq \myexpect{\!}{v^2(t)} = K^2 \frac{G^2 \sigma^2_z + \sigma^2_w}{(1-A^2_c)}.
\end{equation}
In other words, $\mathsf{P}$ is the average transmission power of the controller.
Thus, the \emph{controller-side} signal-to-noise ratio (SNR) for sending the control signal~is 
\begin{equation} \label{snr}
\begin{aligned}
\snr &\triangleq {\mathsf{P}}/{\sigma^2_z} = \frac{\!K^2 G^2 +  K^2 \ssr}{(1-A^2_c)},
\end{aligned}
\end{equation}
where 
$
\ssr \triangleq {\sigma^2_w}/{\sigma^2_z}.
$

%\begin{remark}
%	As the steady-steady state distribution of the plant state $x(t)$ is Gaussian, the control signal of the linear plant is also Gaussian. From~\cite{Gastpar}, our coding-free scheme is the optimal policy in the sense of the long-term performance as both the data source, $v(t)$, and the channel are Gaussian.
%\end{remark}

Given the transmission-power limit of the controller, $\mathsf{P}_0$, the optimization problem is formulated as
\begin{align}
\notag
(\text{S1})\quad &\min_{G,K}\ \myexpect{\!}{x^2(t)}, \\
&\text{s.t.}\  \snr \leq \gamma_0, \text{ and }
 \vert A_c \vert < 1,
\end{align}
where $\gamma_0 \triangleq \mathsf{P}_0/ \sigma^2_z$ is the SNR constraint,
and the first and second constraints are due to the average transmission-power limit of the controller and the closed-loop stability requirement, respectively.
It is clear that (S1) is not a convex problem as $\myexpect{\!}{x^2(t)}$ is not a convex function of $G$ and $K$. We solve (S1) as follows.

First, we analyze the feasibility condition of (S1).
It can be proved that the minimum $\snr$ is equal to 
$
\frac{A^2-1}{H^2},
$
which is achieved when
$
G  K = -\frac{A^2-1}{ A  H}$ and $K \rightarrow 0$,
which also satisfies the stability condition \eqref{Ac}.
Therefore, the feasibility condition of (S1) is 
\begin{equation} \label{feasible}
\frac{\left(A^2-1\right)}{H^2} \leq \gamma_0,
\end{equation}
and we see that there is no feasible solution of (S1), i.e., the plant cannot be stabilized by the remote controller if the channel condition is bad, i.e., a small $\vert H \vert$. Thus, we have the following selection criteria of the plant's control mode.
\begin{criteria*}[Slow-fading-single-plant scenario]
	\normalfont
	If \eqref{feasible} is satisfied, the plant is controlled remotely by the coding-free control method. Otherwise, the plant switches to the self-control mode.
\end{criteria*}

%If the condition \eqref{feasible} is not satisfied, a \emph{control outage} occurs.
%\emph{Note that although the plant cannot be stabilized in the deep fade case (i.e., with very small $\vert H \vert$), by using a frequency diversity technique, we may choose another frequency channel with a better channel quality to control the plant in practice.}
%\myred{As we consider a slow fading channel, according to \eqref{feasible}, the stabilizable probability with the SNR constraint can be written as
%\begin{equation} \label{stable probability}
%P_{\text{stable}} \triangleq \myprobability{H^2 > (A^2-1)/\gamma_0}.
%\end{equation}}

Second, assuming that \eqref{feasible} is satisfied, we solve (S1) in two steps:
\begin{enumerate}
\item Fix $A_c$ such that $ \vert A_c \vert < 1$. We have 	
	\begin{equation}\label{condition}
	G K = \frac{(A_c - A)}{ H}.
	\end{equation}
	%and the optimization problem can be reformulated a convex problem. The optimal cost and the optimal parameters can be obtained straightforwardly.
	Thus, the optimization problem can be written as 
	\begin{align}\label{P1-1-1}
	&(\text{S1-1})\quad\min_{K}\ {\left(\left(\frac{A_c - A}{ H K}\right)^2 \sigma^2_z + \sigma^2_w\right)}\big/{(1-A^2_c)} \\ \label{P1-1-2}
	&\text{s.t.}\  K^2 \leq \frac{1}{\ssr} \left((1-A^2_c) \gamma_0 -(A_c-A)^2/H^2\right).
	\end{align}
	It is easy to see that the optimal cost in \eqref{P1-1-1} is minimized if the equality in \eqref{P1-1-2} holds, and the optimal parameters can be obtained as
	\begin{equation}\label{para_1}
	\left\lbrace
	\begin{aligned}
	J_{\text{ave}}^{\star} &\!= \!\frac{H^2 \gamma_0 \sigma^2_w}{H^2 \gamma_0 (1-A^2_c) -(A_c-A)^2},\\	
	G^\star &\!=\! \frac{1}{K^\star} \frac{A_c-A}{ H},	
	\\
	K^\star &=\! -\!\sqrt{\!\frac{1}{\ssr}\! \left((1-A^2_c) \gamma_0 -(A_c-A)^2/H^2\right)}.
	\end{aligned}
	\right.
	\end{equation}
	
\item Find the optimal $A_c$. Based on Step 1), the problem (S1-1) can be written as
\begin{align}
\label{P1-2-1}
(\text{S1-2})\quad\min_{A_c}&\  \frac{H^2 \gamma_0 \sigma^2_w}{H^2 \gamma_0 (1-A^2_c) -(A_c-A)^2}\\
\notag
\text{s.t.} &\ \vert A_c \vert < 1.
\end{align}
Since the denominator of \eqref{P1-2-1} has a quadratic form, the target function of (S1-2) is minimized when 
\begin{equation}\label{a_star}
A^{\star}_c = \frac{A}{\left(1+ H^2 \gamma_0\right)}.
\end{equation}
Using the feasibility condition \eqref{feasible}, it can be verified that $\vert A^{\star}_c \vert <1$ satisfying the constraint of (S1-2).
\end{enumerate}

Taking $A^{\star}_c$ into \eqref{para_1}, we have the following results.
\begin{theorem} \label{theor:single}
\normalfont
In the slow-fading-single-plant scenario, the optimal cost of the plant and the optimal controller and actuator factors for coding-free control are given as
\begin{equation}
\begin{aligned}
J_{\text{ave}}^\star &= \frac{\sigma^2_w}{1- A^\star_c}= \frac{\sigma^2_w}{1- {A}/{\left(1+ H^2 \gamma_0\right)}},\\
K^\star &= 
-\sqrt{\frac{1}{\ssr}\frac{\gamma_0 \left(H^2 \gamma_0+1 -A^2\right)}{H^2 \gamma_0 +1}},\\
G^\star &=
A H \sqrt{\frac{\gamma_0 \ssr}{{(H^2 \gamma_0 + 1)(H^2 \gamma_0 + 1 -A^2)}}}.
\end{aligned}
\end{equation}
\end{theorem}
\begin{remark}
From Theorem~\ref{theor:single}, the norm of the optimal control  and actuator factors, $\vert K^\star \vert$ and $\vert G^\star \vert$, monotonically increases and decreases with the channel power gain $H^2$, respectively.
\end{remark}

\section{Slow-Fading-Multi-Plant Case: Optimal Design} \label{sec:multi}

\subsection{Joint Controller-Actuator Optimization}
In this section, we consider a slow-fading-multi-plant case and investigate the controller power allocation problem to stabilize the remote plants.
Without loss of generality, we assume that $H_i\geq H_j,\ \forall i<j$.

Based on (S1), the optimization problem to minimize the sum cost of the $M$ plants, is formulated as
\begin{align}
(\text{S2})\quad\min_{\left\lbrace G_i,K_i \right\rbrace}&\ \sum_{i=1}^{M} J_{\text{ave},i} \\
\text{s.t.}&\ \sum_{i=1}^{M} K^2_i \frac{G^2_i  + 
	\ssr}{1-(A+H_i G_i K_i )^2} \leq \gamma_0\\
\label{P21-3}
&\ (A+H_i G_i K_i)^2 < 1,\forall i,
\end{align}
where 
$J_{\text{ave},i} \triangleq \sigma^2_z \frac{G^2_i  + 
	\ssr}{1-(A+H_i G_i K_i)^2}$ is the average cost of plant~$i$. 
Similar to \eqref{feasible}, the feasibility condition of (S2) can be obtained as 
\begin{equation}\label{c2}
\sum_{i=1}^{M} {\left(A^2-1\right)}/{H_i^2} \leq \gamma_0.
\end{equation}
If \eqref{c2} does not hold, the $M$ plants cannot be stabilized simultaneously under the transmission power limit, and we~have the following selection criteria of the plant-control modes.
\begin{criteria*}[Slow-fading-multi-plant scenario]
	\normalfont
The controller selects the set of $M$ plants, i.e., plants $1,2,\cdots M$, for remote coding-free control by the channel coefficients, where $M\leq M_0$ is the largest number of plants that satisfies \eqref{c2}.
The remainder $(M_0-M)$ plants switch to the self-control mode.
\end{criteria*}

%\myred{Similar to the single-plant scenario~\eqref{stable probability}, the stabilizable probability for the multi-plant system can be obtained as
%$
%P_{\text{stable}} \triangleq \myprobability{\sum_{i=1}^{M} {1}/{H_i^2} \leq \frac{\gamma_0}{\left(A^2-1\right)}}.
%$
%}

In the following, we solve problem (S2) by assuming that \eqref{c2} is satisfied with $M>0$. Although (S2) is not a convex problem, 
inspired by the solution of the problem (S1), we first introduce the auxiliary variables $\gamma_{i}$, $i=1,...,M$, which convert (S2) into the following equivalent problem:
\begin{flalign}
(\text{S2'})\quad\min_{\left\lbrace G_i,K_i, \gamma_{i} \right\rbrace}&\ \sum_{i=1}^{M} \sigma^2_z \frac{G^2_i  + 
	\ssr}{1-(A+H_i G_i K_i)^2} \\
\text{s.t.}&\ K^2_i \frac{G^2_i  + 
	\ssr}{1-(A+H_i G_i K_i )^2} \leq \gamma_i, \forall i\\  \label{28}
&\ (A+H_i G_i K_i)^2 < 1,\forall i\\ 
&\ \sum_{i=1}^{M} \gamma_i \leq \gamma_0\\
&\ \gamma_i >0, \forall i.
\end{flalign}
In other words, $\gamma_i$ is the controller's allocated $\snr$ for plant~$i$.
We see that problem (S2') can be converted into $M$ independent problems with the same structure of (S1) when $\{\gamma_i\}$ are given and $(A^2-1)/H^2_i \leq \gamma_i, \forall i$.

Therefore, (S2') can be solved in two steps: 1) solving (S2') with fixed $\gamma_i,\ \forall i$, by using the solution for problem (S1), and 2) optimizing $\{\gamma_i\}$, i.e.,
%\begin{align}
%(\text{S2'-1})\ \min_{ G_i,K_i}&\  J_{\text{ave},i} \\
%\text{s.t.}&\ K^2_i \frac{G^2_i  + 
%	\ssr}{1-(A+H_i G_i K_i )^2} \leq \gamma_i \text{ and } \eqref{28}, 
%\end{align}
%\begin{align}
%(\text{S2'-2})\ \ \min_{\left\lbrace \gamma_i\right\rbrace}&\ \sum_{i=1}^{M} \frac{\sigma^2_w}{1-  {A}/{\left(1+ H^2_i \gamma_i\right)}} \\
%\text{s.t.}&\ \sum_{i=1}^{M} \gamma_i \leq \gamma_0,\\
%& {\left(A^2-1\right)}/{H^2_i} \leq \gamma_i,\forall i.
%\end{align}
\begin{align}
(\text{S2'-1})\ \min_{ G_i,K_i}&\  J_{\text{ave},i} \\
\text{s.t.}&\ K^2_i \frac{G^2_i  + 
	\ssr}{1-(A+H_i G_i K_i )^2} \leq \gamma_i \text{ and } \eqref{28} \\
%\end{align}
%and 
%\begin{align}
(\text{S2'-2})\ \ \min_{\left\lbrace \gamma_i\right\rbrace}&\ \sum_{i=1}^{M} \frac{\sigma^2_w}{1-  {A}/{\left(1+ H^2_i \gamma_i\right)}} \\
\text{s.t.}&\ \sum_{i=1}^{M} \gamma_i \leq \gamma_0,\\
& {\left(A^2-1\right)}/{H^2_i} \leq \gamma_i,\forall i.
\end{align}

Since (S2'-2) is a convex problem, which can be solved by applying KKT conditions, we have the following results.
\begin{theorem} \label{theo:multi}
	\normalfont
In the multi-plant scenario, the optimal sum cost of the plants and the optimal controller and actuator factors are given by, respectively,
\begin{equation}
\begin{aligned}
J_{\text{ave}}^\star &=\sum_{i=1}^{K} \frac{\sigma^2_w}{1-  {A}/{\left(1+ H^2_i \gamma^\star_i\right)}},\\
K^\star_i &= 
-\sqrt{\frac{1}{\ssr}\frac{\gamma^\star_i \left(H^2_i \gamma^\star_i+1 -A^2\right)}{H^2_i \gamma^\star_i +1}},\\
G^\star_i &=
A H_i \sqrt{\frac{\gamma^\star_i \ssr}{{(H^2_i \gamma^\star_i + 1)(H^2_i \gamma^\star_i + 1 -A^2)}}},
\end{aligned}
\end{equation}
where 
\begin{equation}
\gamma^{\star}_i =\max \left\lbrace \frac{\sqrt{A}}{H_i \sqrt{\lambda}} + \frac{A-1}{H^2_i}, \frac{A^2-1}{H^2_i} \right\rbrace,
\end{equation}
and $\lambda>0$ is the unique real root of $\sum_{i=1}^{M} \gamma^{\star}_i = \gamma_0$.
\end{theorem}

\begin{remark}
From Theorem~\ref{theo:multi}, it is interesting to see that the optimal control method allocates the plant with a better channel condition $H_i$, with a lower transmission power, i.e., a smaller $\gamma^{\star}_i$.
{Therefore, the optimal power allocation policy for stabilizing multiple plants is of a channel-inversion type.
Note that although the optimal power allocation policies of the coding-free WNCS and the conventional coding-free communication systems~(see e.g. \cite{AlexW}) are entirely different, these are all channel-inversion type policies.
This is mainly because the two types of problems have distortion-like optimization targets that are the sum of decreasing functions of the channel power gains.}
\end{remark}

To implement the optimal joint controller-actuator design method based on Theorem~\ref{theo:multi}, the controller needs to calculate $M$ pairs of controller-actuator factors and broadcast the actuator factors to the actuators of the the plants selected for coding-free control in the DH. To reduce the communication overhead of the actuator-factor transmissions, in what follows, we propose and optimize two methods: 1) the actuator factors of each plant are  identical and fixed and the controller only optimizes the controller factors and 2) the controller factors of each plant are identical and fixed and each actuator calculates the optimal actuator factor simply based on its own channel coefficient.

\subsection{Optimal Design with Identical Actuator Factors}\label{sec:optimal_K}
We optimize the controller factors $\left\lbrace K_i \right\rbrace$ under the setting that all the actuators have the same fixed actuator factor $G$, while the channel coefficients of the wireless control channels can be different. Thus, we have the following problem after some simplifications:
\begin{align}
\label{P21-1}
(\text{S2-1})\quad\min_{\left\lbrace \tilde{K}_i \right\rbrace}&\ \sum_{i=1}^{M} \frac{\sigma^2_z (G^2+ \ssr)}{1-(A+H_i \tilde{K}_i)^2} \\
\label{P21-2}
\text{s.t.}&\ \sum_{i=1}^{M} \frac{ \tilde{K}^2_i}{1-(A+H_i \tilde{K}_i )^2} \leq \tilde{\gamma}_0, \\ 
\notag
&\text{ and } \eqref{P21-3},
\end{align}
where $\tilde{K}_i \triangleq {K}_i G$ and $\tilde{\gamma}_0 \triangleq \frac{G^2 \gamma_0}{(G^2  + \ssr)}$.

It can be easily proved that the left-hand side of the constraint \eqref{P21-2} is minimized within the stability region \eqref{P21-3} when 
\begin{equation}\label{lhs_constraint}
\tilde{K}_i= -\frac{A^2-1}{A H_i}, \forall i.
\end{equation}
Taking \eqref{lhs_constraint} into \eqref{P21-2}, 
the feasibility condition of (S2-1) can be obtained as 
\begin{equation} \label{c3}
\sum_{i=1}^{M} \frac{A^2-1}{H^2_i} \leq  \frac{G^2 \gamma_0}{(G^2  + \ssr)}.
\end{equation}
If \eqref{c3} does not hold, the $M$ plants cannot be stabilized simultaneously under the transmission power limit, and we~have the following selection criteria of the plant-control modes.
\begin{criteria*}[Slow-fading-multi-plant scenario with identical actuator factors]
	\normalfont
	The controller selects the set of $M$ plants, i.e., plants $1,2,\cdots M$, for remote coding-free control by the channel coefficients, where $M\leq M_0$ is the largest number of plants that satisfies \eqref{c3}.
	The remainder $(M_0-M)$ plants switch to the self-control mode.
\end{criteria*}

\begin{remark}
Comparing this mode selection criteria \eqref{c3} with \eqref{c2} of the joint controller-actuator design problem (S2), although the method of fixed actuator factors has a lower communication payload for parameter exchanging, only a smaller number of plants can be stabilized simultaneously by the controller than the optimal joint-design method.  
\end{remark}

In the following, we solve problem (S2-1) with the assumption that the feasibility condition \eqref{c3} holds.

\begin{figure}[t]
	\renewcommand{\captionfont}{\small} \renewcommand{\captionlabelfont}{\small}	
	\centering
	\includegraphics[scale=1.1]{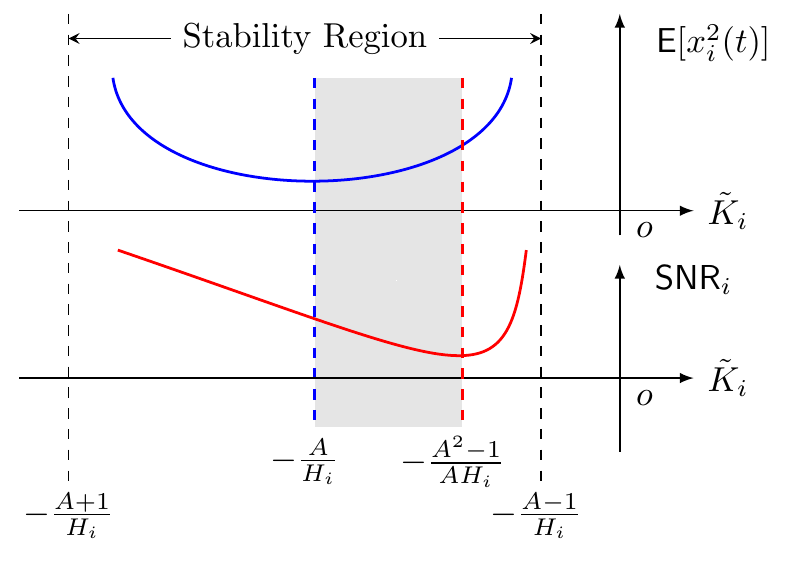}
	\caption{Illustration of $\myexpect{\!}{x^2_i(t)}$ and $\snr_i$ versus $\tilde{K}_i$.}	
	\label{fig:curves}
	%\vspace{-0.5cm}
\end{figure}

First, we analyze the problem (S2-1). Fig.~\ref{fig:curves} illustrates the functions $\myexpect{\!}{x^2_i(t)}$, which is a scaled version of the $i$th term in the target function \eqref{P21-1}, and $\snr_i$, which is a scaled version of the $i$th term in the constraint \eqref{P21-2}, in terms of $\tilde{K}_i$.
It can be proved that both the functions, $\myexpect{\!}{x^2_i(t)}$ and $\snr_i$, are convex in terms of $\tilde{K}_i$ within the stability region \eqref{P21-3}, i.e.,
$
\tilde{K}_i \in \left(-\frac{\left(A+1\right)}{H_i},-\frac{\left(A-1\right)}{H_i}\right),
$
the minimum value of the two functions are achieved at $\tilde{K}_i = -\frac{A}{H_i}$ and $-\frac{A^2-1}{A H_i}$, respectively, 
and the functions are monotonic in the region
$
\tilde{K}_i \in \left[-{A}/{H_i},-{\left(A^2-1\right)}/{\left(A H_i\right)}\right], \ \forall i,
$
as illustrated in Fig.~\ref{fig:curves}.
Therefore, the optimal solution of (S2-1) is within this region (shaded in Fig.~\ref{fig:curves}).
Since $\myexpect{\!}{x^2_i(t)}$ and $\snr_i$ are minimized and maximized, respectively, when $\tilde{K}_i = -\frac{A}{H_i}$,
the summation in \eqref{P21-2} can be maximized as $\sum_{i=1}^{M} \frac{A^2}{H^2_i}$ in the scenario that $\tilde{\gamma}_0 \geq \sum_{i=1}^{M} \frac{A^2}{H^2_i}$, and the optimal design parameters in this scenario are 
\begin{equation} \label{solution1}
\tilde{K}^\star_i = -\frac{A}{H_i}, \forall i.
\end{equation}

Now, we look at the scenario that $\tilde{\gamma}_0 <  \sum_{i=1}^{M} \frac{A^2}{H^2_i}$. In this scenario, it is clear that only the first constraint in \eqref{P21-2} is the active constraint in (S2-1) and the problem is convex.
Using the method of Lagrange multipliers, we can obtain the optimal $\tilde{K}_i$ as
\begin{equation} \label{solution2}
\tilde{K}^\star_i = \frac{D -
	\sqrt{4 A^2 {H^2_i} \lambda' +D^2}}{ 2 A {H_i} \lambda' }, \forall i,
\end{equation}
where $D=(1-A^2)\lambda' +{H^2_i}$ and $\lambda' > 0$ is the unique real root of 
$
\sum_{i=1}^{M} \frac{ (\tilde{K}^\star_i)^2}{1-(A+H_i \tilde{K}^\star_i )^2} = \tilde{\gamma}_0.
$

\begin{proposition}\label{prop:K}
	\normalfont
When the actuator factors are fixed, i.e., $G_i=G, \forall i\in\{1,\cdots,M\}$,	the optimal control factor of actuator~$i$ is given as
\begin{equation}
\tilde{K}^\star_i = \begin{cases}
-\frac{A}{H_i} & \tilde{\gamma}_0 \geq \sum_{i=1}^{M} \frac{A^2}{H^2_i}\\
\frac{D -
	\sqrt{4 A^2 {H^2_i} \lambda' +D^2}}{ 2 A {H_i} \lambda' }, &\text{otherwise},
\end{cases}
\end{equation}
where $D$ and $\lambda'$ are defined under \eqref{solution2}.
\end{proposition}

\begin{remark}
	From Proposition~\ref{prop:K}, it can be proved that the norm of the controller factor, $\vert \tilde{K}^\star_i \vert$, decreases with the increasing of $\vert H_i \vert $. 
	In other words, a larger controller factor is applied to control the plant with a smaller channel power gain.
\end{remark}

\subsection{Optimal Design with Identical Controller Factors} \label{sec:optimal_G}
%Similarly, we can investigate the optimal actuator factors $\left\lbrace G_i \right\rbrace$ under the setting that the controller applies the same controller factor $K$ for each plant.
%It can be proved that both the optimal $\vert {G}_i \vert$ and $\snr_i$ decrease with $\vert H_i \vert $, i.e., a channel-inversion-like power allocation policy.
%The detailed analysis is omitted due to the space limitation. 
Now we investigate the optimal actuator factors $\left\lbrace G_i \right\rbrace$ under the setting that the controller applies the same fixed controller factor $K$ for each plant. Thus, we have the following problem after some simplifications:
\begin{align}
\notag
(\text{S2-2})\quad\min_{\left\lbrace G_i \right\rbrace}&\ J_{\text{ave}} = \sum_{i=1}^{M} \sigma^2_z \frac{G^2_i +\ssr}{1-(A+H_i K G_i)^2} \\
\notag
\text{s.t.}&\ J_{\text{ave}} \leq \frac{\sigma^2_z {\gamma}_0}{K^2} \text{ and }\eqref{P21-3},
\end{align}
which is a convex problem. The optimal parameter is obtained~as 
\begin{equation} \label{optimal_solution}
G^\star_i = \frac{E_i - \sqrt{4 A^2 H^2_i K^2 \ssr + E_i^2}}{2 A H_i K}, \forall i,
\end{equation}
where
\begin{equation}
E_i =  \left(1-A^2 + H_i^2 K^2 \ssr\right),
\end{equation}
and the feasibility condition is 
\begin{equation} \label{G_condition}
J^\star_{\text{ave}} \leq \sigma^2_w/K^2.
\end{equation}
Therefore, we have the following results.
\begin{criteria*}[Slow-fading-multi-plant scenario with identical controller factors]
	\normalfont
	The controller selects the set of $M$ plants, i.e., plants $1,2,\cdots M$, for remote coding-free control by the channel coefficients, where $M\leq M_0$ is the largest number of plants that satisfies \eqref{G_condition}.
	The remainder $(M_0-M)$ plants switch to the self-control mode.
\end{criteria*}

\begin{proposition}\label{prop:G}
	\normalfont
	When the controller factors are fixed, i.e., $K_i=K, \forall i\in\{1,\cdots,M\}$, the optimal actuator factor~$i$ is given in \eqref{optimal_solution}.
\end{proposition}

\begin{remark}
	From Proposition~\ref{prop:G}, it can be proved that $\vert \tilde{G}^\star_i \vert$ decreases with $\vert H_i \vert $. 
	In other words, a larger actuator factor is applied to control the plant with a smaller channel coefficient.
\end{remark}
%\begin{remark}
%	From the optimal solution \eqref{optimal_solution}, it can be proved that both $\vert \tilde{G}^\star_i \vert$ and $\snr_i$ decreases with $\vert H_i \vert $. 
%	In other words, a larger actuator factor and a larger transmit power are applied to control the plant with a smaller channel coefficient, which is also a channel-inversion type policy.
%\end{remark}

\section{Fast-Fading Case: Analysis and Optimization}\label{sec:fast}
%In this section, we consider the optimal design problem of the single-plant case, and hence drop out the plant index $i$ in the following analysis.

In the fast-fading scenario, it is not practical to have full channel-state information (CSI) at the controller nor the actuators, since the channel-estimation time for precise CSI is long and comparable to the channel-coherence time. Thus,
we investigate two cases: 
\begin{enumerate}
	\item Control without CSI. The CSI is unknown to neither the controller nor the actuators.
	\item Control with partial CSI. Only the signs of the real channel coefficients are known to the controller, and the actuators do not have any CSI.\footnote{Although the channel estimation method can be the same as the conventional pilot-based one~\cite{BOOKTse}, to estimate the sign of the real channel, i.e., a binary detection process, the required phase for channel estimation is much shorter than that of the channel coefficient. Also, the effect of the the sign-estimation error will be taken into account in our future work.}
\end{enumerate}

\subsection{Single-Plant Case}
{Similar to Sec.~\ref{sec:single}, we drop out the plant index $i$ in the following analysis of the single-plant case.}

Taking \eqref{v} and \eqref{u} into \eqref{P}, the dynamic process of the plant state in the fast-fading case can be rewritten as
\begin{equation} \label{x-x2}
\begin{aligned}
x(t+1) 
&= A_c(t) x(t) +  G z(t) + w(t),
\end{aligned}
\end{equation}
where
$A_c(t) \triangleq A +  G K H(t)$.
Since $H(t)$, $x(t)$, $z(t)$ and $w(t)$ are independent of each other, we have
\begin{equation} \label{Ex2}
\myexpect{\!}{x^2(t+1)}=\myexpect{\!}{A^2_c(t)} \myexpect{\!}{x^2(t)} + G^2 \sigma^2_z + \sigma^2_w.
\end{equation}
To make the long-term average cost $J_\text{ave}$ in \eqref{def:cost_longterm} bounded, $\myexpect{\!}{x^2(t)}$ should be bounded when $t\rightarrow \infty$.
From \eqref{Ex2}, it is clear that $\myexpect{\!}{x^2(t)}$ is bounded iff $\myexpect{\!}{A^2_c(t)}<1$. Thus, the stability condition is obtained as below.
\begin{definition}[Stability condition in fast-fading channel]
	\normalfont
	The plant \eqref{P} is closed-loop  mean-squared stable using the coding-free control method \eqref{v} and \eqref{u} iff 
	\begin{equation} \label{Ac2}
	\myexpect{\!}{A^2_c(t)}<1.
	\end{equation}
\end{definition}

\subsubsection{Control without CSI}
Since $H(t)\sim \mathcal{N}(0,\sigma^2_h)$, we have
\begin{equation}
\begin{aligned}
\myexpect{\!}{A^2_c(t)} 
&=  \int_{-\infty}^{\infty}\left(A+GK H\right)^2 \exp\left(- \frac{x^2}{2 \sigma^2_h}\right) \frac{1}{\sqrt{2\pi \sigma^2_h}} \mathrm{d}H\\
&= A^2+ (GK)^2 \sigma^2_h>1.
\end{aligned}
\end{equation}
Therefore, from the updating rule of the plant-state covariance \eqref{Ex2}, $\myexpect{\!}{x^2(t)}$ exponentially increases with $t$, and thus the long-term average cost function \eqref{def:cost_longterm} cannot be bounded.
\begin{remark}
	In the fast-fading scenario, the plant cannot be stabilized by the remote controller using the coding-free method in \eqref{v} and \eqref{u}, if the CSI is unknown to neither the controller nor the actuators.
\end{remark}
In general, making $\vert A_c(t)\vert$ less than one with a certain high probability  is crucial for stabilizing a remote plant. As $A$ is greater than one, $\vert A_c(t)\vert$ is absolutely greater than one if the sign of $GKH(t)$ is positive, while it may be less than one if the sign of $GKH(t)$ is negative. 
Thus, it is important to properly control the sign of $GKH(t)$.
Therefore, in the following, we investigate whether the plant is stabilizable with a minimum available CSI, i.e., only the signs of the real channel coefficients are known to the controller.

\subsubsection{Control with Partial CSI} \label{sec:partialCSI} 
Since the controller knows the sign of the channel coefficient, the controller factor can be designed as a function of the sign of the channel coefficient.
Therefore, we consider the following coding-free control method with partial CSI as
\begin{equation} \label{new_method}
\begin{cases}
G(t)=G\\
K(t) = \text{sgn}(H(t)) K,
\end{cases}
\end{equation}
where $\text{sgn}(\cdot)$ is the signum function.
In other words, $G(t)$ and $\vert K(t)\vert$ are fixed and do not change with time, where $G>0$ and $K<0$ and are the same as the slow-fading case, while the sign of the controller factor, i.e., $K(t)$, changes with time and depends on the sign of the channel coefficient. In this way, the term $G(t)K(t)H(t)=GK\vert H(t)\vert$ is always non-positive.

Thus, we have
\begin{equation}\label{new_Ac}
\begin{aligned}
\myexpect{\!}{A^2_c(t)} 
&= 2 \int_{0}^{\infty}\left(A+GK H\right)^2 \exp\left(- \frac{x^2}{2 \sigma^2_h}\right) \frac{1}{\sqrt{2\pi \sigma^2_h}} \mathrm{d}H\\
&=\sigma^2_h (GK)^2 + 2 \sqrt{\frac{2}{\pi } \sigma^2_h} A (GK) + A^2,
\end{aligned}
\end{equation}
which is a quadratic function of $GK$. As $\myexpect{\!}{A^2_c(t)}$ is independent of $t$, its time index is dropped out in the following.
When $GK = - A \sqrt{\frac{2}{\pi  \sigma^2_h}} $, $\myexpect{\!}{A^2_c} $ is minimized and equal to $A^2\left(1 - \frac{2}{\pi}\right)$. If $\myexpect{\!}{A^2_c}\geq 1$, the plant cannot be stabilized as discussed earlier.
Then, we can easily obtain the following results.
\begin{proposition} \label{prop:fast_stability}
	\normalfont
The plant cannot be stabilized by the remote controller using the coding-free control method~\eqref{new_method} with partial CSI, if 
\begin{equation}
	A^2 \eta \geq 1,
\end{equation}
where 
\begin{equation}
	\eta = 1-\frac{2}{\pi} \approx 0.3634.
\end{equation}
\end{proposition}

From Proposition~\ref{prop:fast_stability}, if $\vert A \vert \geq \frac{1}{\sqrt{\eta}} \approx 1.65$, the plant cannot be stabilized no matter how large the controller's transmission power is. In practice, $\vert A \vert $ is small and usually no larger than $1.5$ (see e.g. \cite{TCP,Randomdelay,Gatsis}). Thus, we assume that $\vert A \vert  < \frac{1}{\sqrt{\eta}}$ in the following analysis.

%\begin{lemma}
%$\myexpect{\!}{A^2_c} <1$ iff
%$\left(-\sqrt{\frac{2 A^2}{\pi \sigma^2_h}}-\sqrt{\frac{\pi- A^2(\pi-2)}{\pi \sigma^2_h}},-\sqrt{\frac{2 A^2}{\pi \sigma^2_h}}+\sqrt{\frac{\pi- A^2(\pi-2)}{\pi \sigma^2_h}}\right)$
%\end{lemma}

Using the updating rule of the state variance \eqref{Ex2}, we further have
\begin{equation} \label{x-02}
\begin{aligned}
\myexpect{\!}{x^2(t)}&= \left(\myexpect{\!}{A^2_c}\right)^t \myexpect{\!}{x^2(0)}\\ 
&+ \sum_{j=0}^{t-1} \left(\myexpect{\!}{A^2_c}\right)^{t-1-j} \left(G^2 \sigma^2_z + \sigma^2_w\right).
\end{aligned}
\end{equation}
As we consider the long-term performance of the closed-loop stable dynamic process, 
letting $t \rightarrow \infty$ and using the stability condition that $\myexpect{\!}{A^2_c}<1$, we have 
\begin{equation} \label{average_x2_2}
J_{\text{ave}} = \lim\limits_{t\rightarrow \infty}\myexpect{\!}{x^2(t)} = \frac{G^2 \sigma^2_z + \sigma^2_w}{1-\myexpect{\!}{A^2_c}}
%= \frac{G^2 \sigma^2_z + \sigma^2_w}{1-\left(\sigma^2_h (GK)^2 + 2 \sqrt{\frac{2}{\pi } \sigma^2_h} A (GK) + A^2\right)}.
\end{equation}
where $\myexpect{\!}{A^2_c}$ is given in \eqref{new_Ac}.
From \eqref{v} and \eqref{average_x2_2}, the average transmission power of the controller~is
\begin{equation}
\mathsf{P}  = K^2 \frac{G^2 \sigma^2_z + \sigma^2_w}{(1-\myexpect{\!}{A^2_c})}.
\end{equation}

Similar to the slow-fading scenario, given the transmission-power limit of the controller, $\mathsf{P}_0$, the optimization problem is formulated as
\begin{align}
\notag
(\text{F1})\quad &\min_{G,K}\ J_{\text{ave}} = \frac{G^2 \sigma^2_z + \sigma^2_w}{(1-\myexpect{\!}{A^2_c})}, \\ \label{fast_snr}
&\text{s.t.}\  \snr = K^2 \frac{G^2  + \ssr}{(1-\myexpect{\!}{A^2_c})} \leq \gamma_0,\\ \label{fast_Ac}
&\hspace{0.7cm}\myexpect{\!}{A^2_c} < 1.
\end{align}
Similar to problem (S1), $\snr$ in \eqref{fast_snr} is minimized as
\begin{equation}
\snr=
\frac{A^2-1}{(1-A^2 \eta) \sigma^2_h}
\end{equation}
when $K\to 0$ and 
\begin{equation}
GK=-\frac{A^2-1}{A \sqrt{\frac{2 \sigma^2_h}{\pi }}}.
\end{equation}
To satisfy both the constraints \eqref{fast_snr} and \eqref{fast_Ac}, we can obtain the following mode-selection criteria.

\begin{criteria*}[Fast-fading-single-plant scenario]
	\normalfont
The plant is controlled remotely by the coding-free control method if 
\begin{equation}
	A^2 \leq \frac{1+\sigma^2_h \gamma_0}{1+\eta\sigma^2_h \gamma_0}.
\end{equation}
Otherwise, the plant switches to the self-control mode.
\end{criteria*}

Although (F1) is not a convex problem, similar to problem (S1), we solve (F1) by fixing $GK$ first and solving the optimal $G$ and $K$, and then solving the optimal $GK$. We can obtain the following results.

%\begin{theorem}
%	\begin{equation}
%		\begin{aligned}
%	J_{\text{ave}}^{\star} &= \frac{\gamma_0 \sigma^2_w}{(1-\myexpect{\!}{A^2_c} ) \gamma_0-U^2},\\	
%G^\star &= \frac{U}{K^\star},	
%\\
%K^\star &= - \sqrt{\frac{1}{\ssr} \left((1-\myexpect{\!}{A^2_c}) \gamma_0 -U^2\right)}.
%		\end{aligned}
%	\end{equation}
%\end{theorem}

%
%\begin{equation}
%U^\star = -\frac{\sqrt{\frac{2}{\pi } \sigma^2_h} A \gamma_0}{1+  \sigma^2_h \gamma_0}
%\end{equation}
%
%\begin{equation}
%\begin{aligned}
%J_{\text{ave}}^{\star} &= \frac{\pi  \sigma^2_h (\gamma_0 \sigma^2_h+1)}{2 A^2 \gamma_0 \sigma^2_h -\pi  \left(A^2-1\right) (\gamma_0 \sigma^2_h+1)}\\
%&= \frac{\pi  \sigma^2_h (\gamma_0 \sigma^2_h+1)}{A^2 (2 \gamma_0 \sigma^2_h-\pi  (\gamma_0 \sigma^2_h+1))+\pi  (\gamma_0 \sigma^2_h+1)}
%\end{aligned}
%\end{equation}

\begin{theorem} \label{theor:single_2}
	\normalfont
	In the fast-fading-single-plant scenario, the optimal cost of the plant and the optimal controller and actuator factors for coding-free control are given as
	\begin{equation}
		\begin{aligned}
		J_{\text{ave}}^{\star} &= \frac{\gamma_0 \sigma^2_w}{(1-\myexpect{\!}{A^2_c}^\star ) \gamma_0-(U^{\star})^2}\\	
		G^\star &= \frac{U^\star}{K^\star}	
		\\
		K^\star &= - \sqrt{\frac{1}{\ssr} \left((1-\myexpect{\!}{A^2_c}^\star) \gamma_0 -(U^{\star})^2\right)}
		\end{aligned}
	\end{equation}
	where 
	\begin{align}
	 &U^{\star}=  -\frac{\sqrt{\frac{2}{\pi } \sigma^2_h} A \gamma_0}{1+  \sigma^2_h \gamma_0}\\
	 &\myexpect{\!}{A^2_c}^\star=\sigma^2_h (U^{\star})^2 + 2 \sqrt{\frac{2}{\pi } \sigma^2_h} A U^{\star} + A^2,
	\end{align}
\end{theorem}
%\begin{remark}
%	From Theorem~\ref{theor:single_2}, the norm of the optimal control  and actuator factors, $\vert K^\star \vert$ and $\vert G^\star \vert$, monotonically increases and decreases with the variance of the channel coefficient $\sigma^2_h$, respectively.
%\end{remark}

\subsection{Multi-Plant Case}
Now, we investigate the optimal coding-free control for 
multiple plants in fast-fading channels with partial CSI.
We have the following optimization problem
\begin{align}
(\text{F2})\quad\min_{\left\lbrace G_i,K_i \right\rbrace}&\ \sum_{i=1}^{M} J_{\text{ave},i} \\
\text{s.t.}&\ \sum_{i=1}^{M} K^2 \frac{G^2  + \ssr}{(1-\myexpect{\!}{A^2_{c,i}})} \leq \gamma_0,\\
&\hspace{0.7cm}\myexpect{\!}{A^2_{c,i}}  \leq 1 ,\forall i,
\end{align}
where 
$J_{\text{ave},i} \triangleq  \frac{G^2 \sigma^2_z + \sigma^2_w}{1-\myexpect{\!}{A^2_{c,i}}}$ is the average cost of plant~$i$, {$\myexpect{\!}{A^2_{c,i}} \triangleq \sigma^2_{h,i} (G_i K_i)^2 + 2 \sqrt{\frac{2}{\pi } \sigma^2_{h,i}} A (G_i K_i) + A^2$}, and $\sigma^2_{h,i}\triangleq \myexpect{\!}{H^2_{i}}$ is the average channel power gain of channel $i$.
Without loss of generality, we assume that $\sigma^2_{h,i}\geq \sigma^2_{h,j},\ \forall i<j$.

Similar to the slow-fading case, the mode-selection criteria is given as
\begin{criteria*}[Fast-fading-multi-plant scenario]
	\normalfont
	The controller selects the set of $M$ plants, i.e., plants $1,2,\cdots M$, for remote coding-free control by the average channel power gains, where $M\leq M_0$ is the largest number of plants that satisfies 
	\begin{equation}
\sum_{i=1}^{M}	\frac{A^2-1}{(1-A^2 \eta) \sigma^2_{h,i}} \leq \gamma_0.
	\end{equation}
	The remainder $(M_0-M)$ plants switch to the self-control mode.
\end{criteria*}

Following the same steps as (S2), we have the following results.
\begin{theorem} \label{theo:multi2}
	\normalfont
	In the fast-fading-multi-plant scenario, the optimal sum cost of the plants and the optimal controller and actuator factors are given by, respectively,
	\begin{equation}
	\begin{aligned}
		J_{\text{ave}}^{\star} &= \sum_{i=1}^{M} \frac{\gamma^{\star}_i \sigma^2_w}{(1-\myexpect{\!}{A^2_{c,i}}^\star ) \gamma^{\star}_i-(U^{\star}_i)^2}\\	
G^\star_i &= \frac{U^\star_i}{K^\star_i}	
\\
K^\star_i &= - \sqrt{\frac{1}{\ssr} \left((1-\myexpect{\!}{A^2_{c,i}}^\star) \gamma^{\star}_i -(U^{\star}_i)^2\right)}
	\end{aligned}
	\end{equation}
	where $	U^{\star}_i =  -\frac{\sqrt{\frac{2}{\pi } \sigma^2_{h,i}} A \gamma^{\star}_i}{1+  \sigma^2_{h,i} \gamma^{\star}_i}$, $\myexpect{\!}{A^2_{c,i}}^\star=\sigma^2_h (U^{\star}_i)^2 + 2 \sqrt{\frac{2}{\pi } \sigma^2_h} A U^{\star}_i + A^2$ and 
	\begin{equation}
	\gamma^{\star}_i =\frac{  \left(A^2-1\right)+ \sqrt{\frac{2}{\pi} } \frac{ A \sigma^2_{h,i}}{\sqrt{\lambda'' }}}{\left(1 -\eta  A^2\right) \sigma^2_{h,i}},
	\end{equation}
	and $\lambda''>0$ is the unique real root of $\sum_{i=1}^{M} \gamma^{\star}_i = \gamma_0$.
\end{theorem}

%
%\begin{equation}
%\gamma_i= \frac{\pi  \left(A^2-1\right)+\frac{\sqrt{2 \pi } A \sigma^2_{h,i}}{\sqrt{\lambda }}}{\left(\pi -(\pi -2) A^2\right) \sigma^2_{h,i}}
%\end{equation}

\section{Numerical Results} \label{sec:num}
In this section, we numerically present the system performance of the proposed coding-free control method for single- and multi-plant cases.
Unless otherwise stated, we set $A = 1.5$, $\sigma^2_w = 0.1$~\cite{TCP}, the noise power at the actuator $\sigma^2_z =-40$~dBm, the controller's transmission-power limit $\mathsf{P}_0 = 20$~dBm, the number of control symbols in a control process $T=500$.

In Sections \ref{sec:num_process}, \ref{sec:num_dig} and \ref{sec:num_slow}, we present results for the \emph{slow-fading case}. We assume Rayleigh block fading channel for evaluating the system performance that averages over multiple control processes. 
%We set the average channel power gain as $10^{-4}$.
For the coding-free control method, we use the optimal controller and actuator factors in Theorems~\ref{theor:single} and~\ref{theo:multi}.

In Section~\ref{sec:num_fast}, we  present the results for the \emph{fast-fading case}, and assume that the channel varies symbol by symbol and follows a Gaussian distribution, and only the sign of the channel coefficients are known to the controller. We use the coding-free control method presented in Sec.~\ref{sec:partialCSI} and apply the optimal controller and actuator factors in Theorems~\ref{theor:single_2} and~\ref{theo:multi2}.

\subsection{State Dynamics of a Single Plant} \label{sec:num_process}
\begin{figure}[t]
	\renewcommand{\captionfont}{\small} \renewcommand{\captionlabelfont}{\small}
	\centering
%	\vspace{-0.6cm}	
	\includegraphics[scale=0.63]{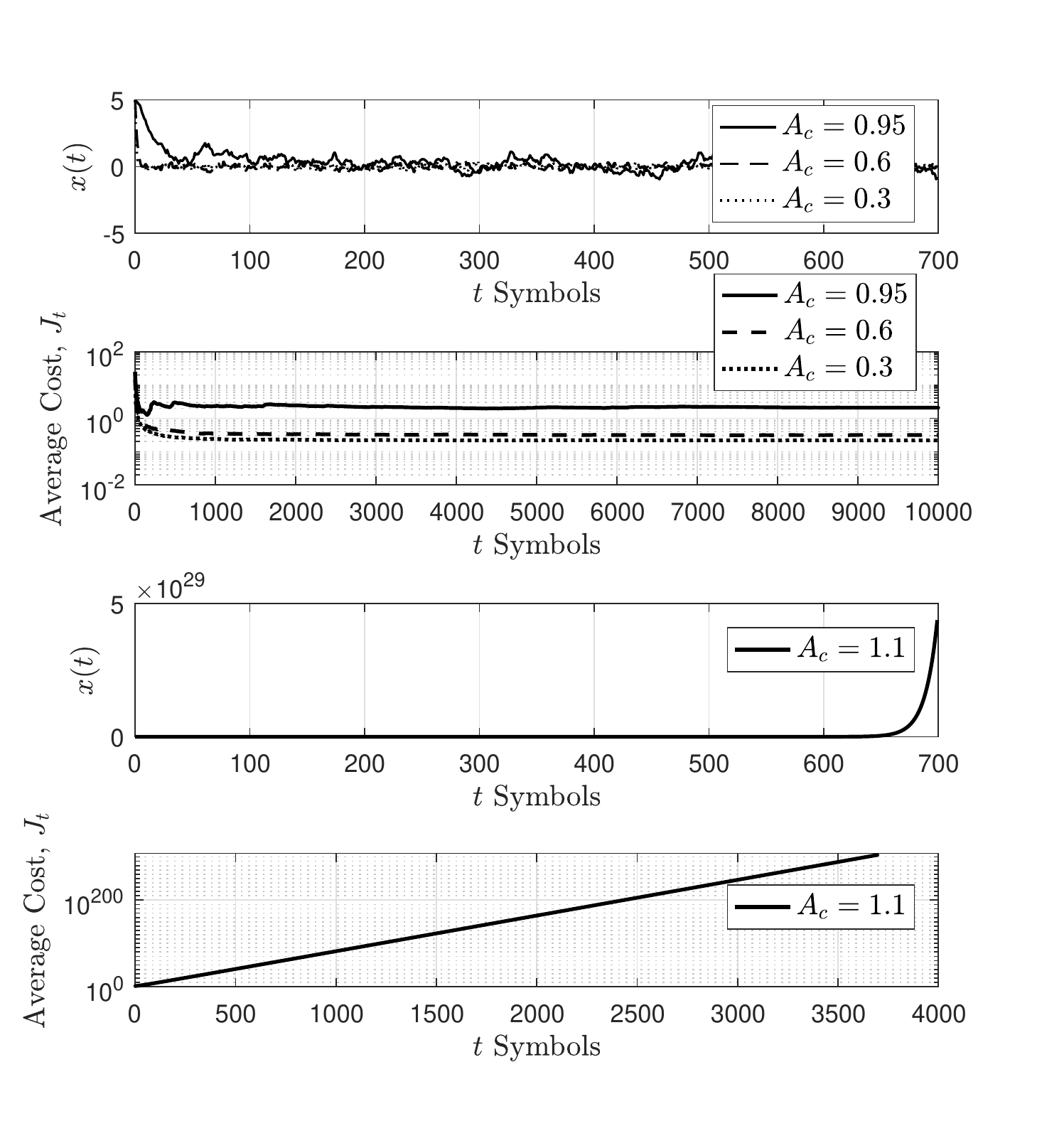}
	\vspace{-0.5cm}	
	\caption{The plant state $x(t)$ and the average cost function in term of the number of control symbols with different closed-loop parameter~$A_c$.}
	\label{fig:decay_process}
%	\vspace{-0.5cm}
\end{figure}

In Fig.~\ref{fig:decay_process}, the plant state and the average cost of a single plant with different closed-loop plant parameter $A_c$ are plotted using~\eqref{x-x} and~\eqref{def:cost}, respectively, where the initial state $x(0) = 5$. 
In the stable scenario, i.e., $A_c < 1$,
we see that the plant state $x(t)$ can be driven close to the steady state with $100$ control symbols, and the average cost $J_t$ in \eqref{def:cost} with $t$ control symbols is quite close to the long-term average cost $J_{\text{ave}}$ as long as $t \geq 500$. \emph{Therefore, it is a good choice to use the long-term cost as the system performance metric.}
In the unstable scenario, i.e., $A_c > 1$ and the stability condition~\eqref{def:cost_longterm} is not satisfied, we see that the average cost is too large to be accepted when $t > 100$. \emph{Therefore, the long-term stability condition is still useful even when we consider a control process with a finite horizon.}

%\begin{figure}[t]
%	\renewcommand{\captionfont}{\small} \renewcommand{\captionlabelfont}{\small}
%	\centering
%	\includegraphics[scale=0.55]{decay_cost.pdf}
%	%	\vspace{-0.9cm}	
%	\caption{The average cost function in term of the number of control symbols with different closed loop parameter $A_c$.}
%	\label{fig:decay_cost}
%	\vspace{-0.5cm}
%\end{figure}

\subsection{Performance Comparison: Coding-Free Control v.s. Coding-Based Control} \label{sec:num_dig}
\begin{figure}[t]
	\renewcommand{\captionfont}{\small} \renewcommand{\captionlabelfont}{\small}
	\centering
	%	\vspace{-0.6cm}	
	\includegraphics[scale=0.78]{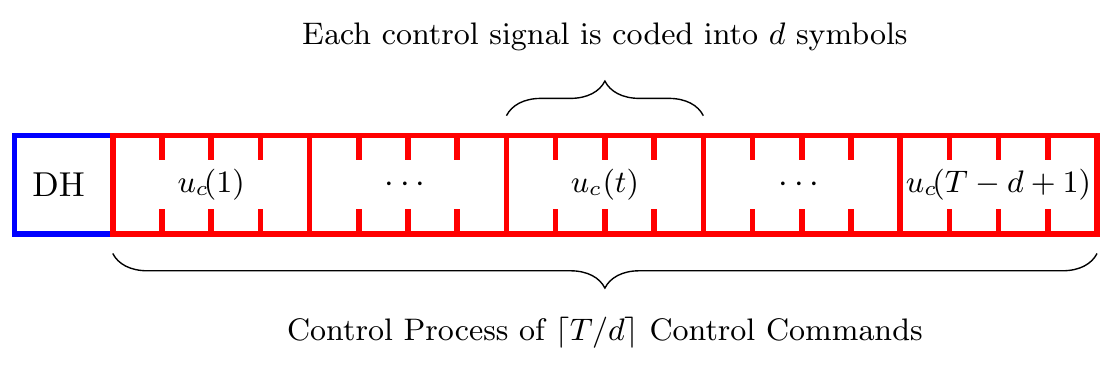}
	%	\vspace{-0.9cm}	
	\caption{Coding-based control protocol.}
	\label{fig:process_digital}
	%	\vspace{-0.5cm}
\end{figure}

We consider the single-plant case for performance comparison.
For coding-based control methods, we adopt Bose-Chaudhuri-Hocquenghem (BCH) codes and quadrature amplitude modulation (QAM) schemes, where the transmissions utilizes two degrees of freedom of the complex channels.
We use the maximum-likelihood method for signal detection and decoding.
To be specific, during the control process of $T$ transmission symbols, the controller generates control signals every $d$ symbols, where $d \ll T$, as illustrated in Fig.~\ref{fig:process_digital}. The control signal generated at time $t$, $u_c(t)$, is first quantized into $K$ bits and then converted into a $N$-bit sequence using the BCH $(N,K)$ code. Last, the bit sequence is modulated into $d \triangleq \lceil N/L \rceil$ symbols using the $2^L$-QAM scheme.\footnote{Note that it is still an open problem in the literature to optimally design the parameters $\{N,K,L\}$.} 
Thus, the quantized control signal will be coded and transmitted to the actuator in $d$ symbols, and \emph{$d$ is the transmission latency of the conventional coding-based control method.}

In this scenario, the optimal control signal before quantization is~\cite{dead}% $u_c(t) = -A^d x(t) $ 
\begin{equation}
u_c(t) = \begin{cases}
-A^d x(t) , &t =1,1+d,1+2d,\cdots\\
0, &\text{otherwise}.
\end{cases}
\end{equation}
Once received and decoded correctly, the detected control signal is implemented by the actuator as in \eqref{P};
if the detection fails, the actuator does not control the plant~\cite{TCP},~i.e.,
\begin{equation}
u(t) = \begin{cases}
\tilde{u}_c(t-d+1),& \text{if successful detection at } \\
&\qquad\qquad\qquad t=1+k d, k\in \mathbb{N} \\
0 & \text{otherwise},
\end{cases}
\end{equation}
where $\tilde{u}_c(t-d+1)$ denotes the quantized control signal of $u_c(t-d+1)$, and $\mathbb{N}$ denotes the set of non-negative integers.
Here, we only focus on the effect of the transmission latency and reliability of the coding-based control on the system performance, and thus, we assume that the quantizer does not introduce any distortion of the original control signal~\cite{TCP}, i.e., $u_c(t) = \tilde{u}_c(t)$.
That is to say what we obtain is, in fact, an upper bound on the performance of the coding-based control method. 

\begin{figure}[t]
	\renewcommand{\captionfont}{\small} \renewcommand{\captionlabelfont}{\small}
	\centering
%	\vspace{-0.3cm}	
	\includegraphics[scale=0.63]{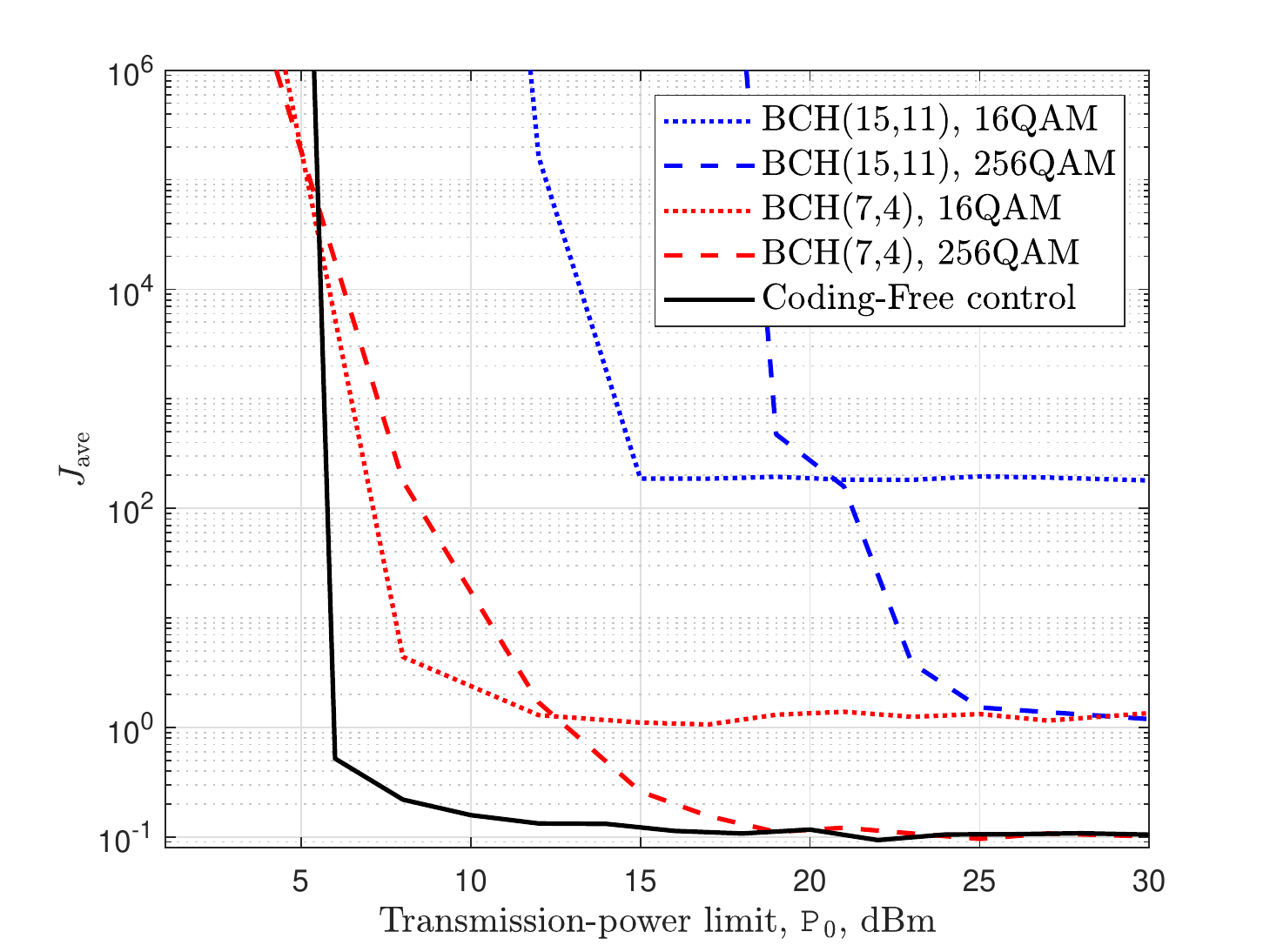}
%	\vspace{-0.5cm}	
	\caption{The average cost of a single plant with the coding-free and coding-based control methods.}
	\label{fig:coding}
%	\vspace{-0.2cm}
\end{figure}

In Fig.~\ref{fig:coding}, the average costs of a single plant with coding-free and practical coding-based control methods and different transmission-power limits are plotted, where the average cost $J_{\text{ave}}$ is simulated by running a control process with $500$ symbols and taking average using \eqref{def:cost}, and the channel coefficient is $H = 0.01$.
The methods BCH(15,11)-16QAM and BCH(7,4)-256QAM have the transmission latency of $4$ symbols and $1$ symbol, respectively, and the methods BCH(15,11)-256QAM and BCH(7,4)-16QAM have the same transmission latency, i.e., $2$ symbols.
In the high SNR regime, i.e., $\mathsf{P}_0 > 20$~dBm, \emph{we see that the average costs of the plant with different coding-based control methods simply depend on the transmission latency}, and the coding-free method almost achieves the lowest average cost achieved by the coding-based method, i.e., BCH(7,4)-256QAM. 
Also, we see that the coding-free method is optimal in the practical range of transmission power, e.g., $6-20$~dBm.
In the low SNR regime, the coding-free method results in an infinite cost of the plant with the power limit, however, the coding-based methods, i.e., BCH(7,4)-16QAM and BCH(7,4)-256QAM, have bounded (yet large) average costs. This implies that coding-based methods can be superior when the SNR is low.

\subsection{Optimal Coding-Free Control of Multi-Plant Systems in Slow-Fading Channels} \label{sec:num_slow}
\begin{figure}[t]
	\renewcommand{\captionfont}{\small} \renewcommand{\captionlabelfont}{\small}
	\centering
%	\vspace{-0.3cm}	
	\includegraphics[scale=0.63]{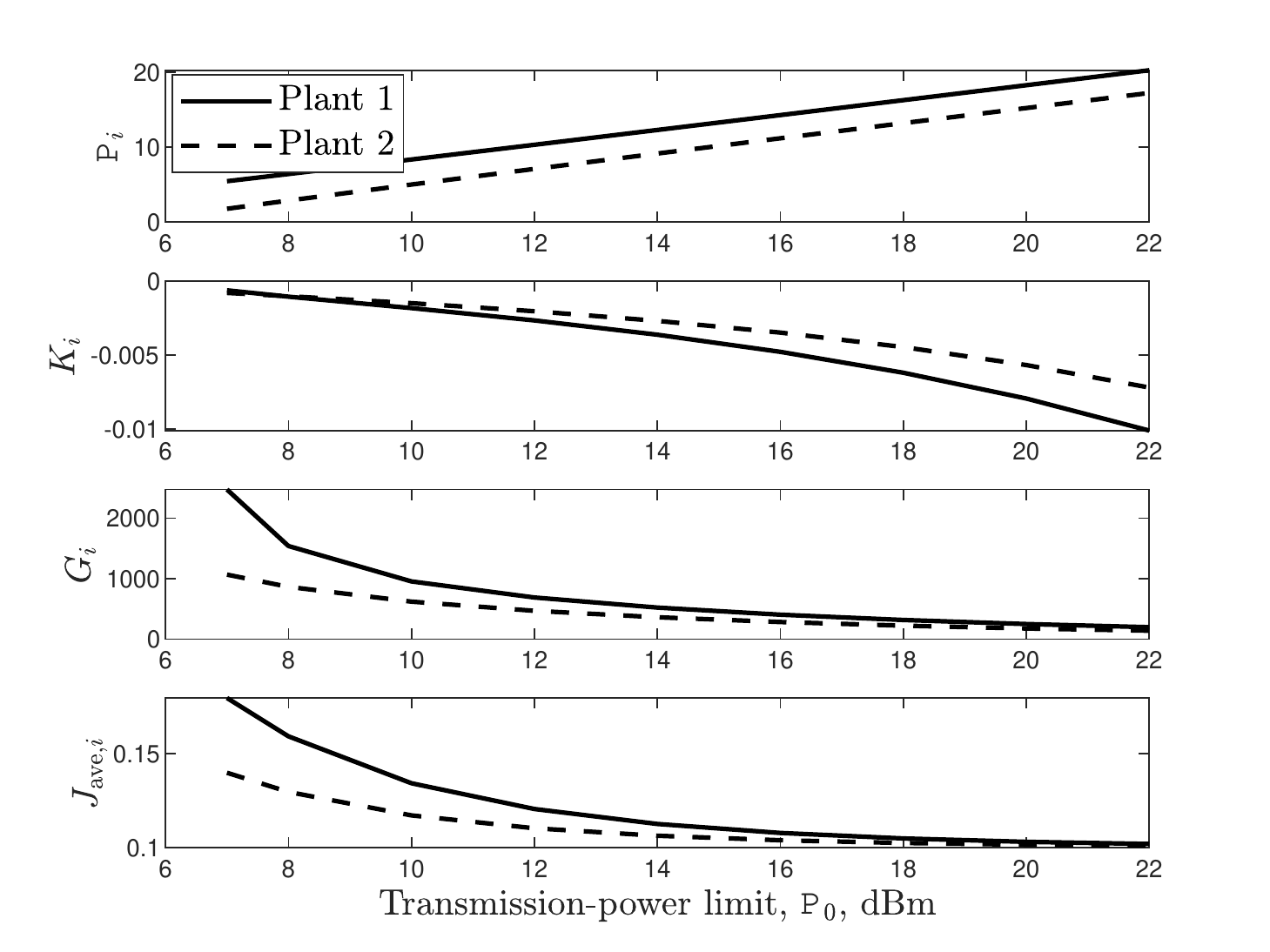}
%	\vspace{-0.5cm}	
	\caption{The optimal control parameters and costs of the plants of the coding-free control method with different transmission-power limit in slow-fading channels.}
	\label{fig:optimal_parameter}
%	\vspace{-0.3cm}
\end{figure}

In Fig.~\ref{fig:optimal_parameter}, we plot the optimal allocated transmission power $\mathsf{P}_i$, control factor $K_i$, actuator factor $G_i$ and the average costs of a two-plant system of the optimal coding-free control policy, where $H_1= 0.01$ and $H_2 =0.02$.
We see that it is not possible to stabilize both the plants when the transmission-power limit, $\mathsf{P}_0$, is less than $6$~dBm.
When $\mathsf{P}_0>6$~dBm, the optimal allocated power to each plant increases, while the optimal controller and actuator factors and the average cost of each plant decreases with the increasing of the transmission-power limit $\mathsf{P}_0$.
Also, we see that although the plant with a worse channel condition, i.e., plant~1, is allocated with a larger transmission power and has a larger actuator factor than that of plant~2, the average cost of plant~1 is still larger than plant~2.
Also, we see that the controller factor of plant 1 is larger and smaller than plant 2 in the low (e.g., $\mathsf{P}_0<8$~dBm) and high SNR regimes, respectively.

\begin{figure}[t]
	\renewcommand{\captionfont}{\small} \renewcommand{\captionlabelfont}{\small}
	\centering
	%	\vspace{-0.3cm}	
	\includegraphics[scale=0.63]{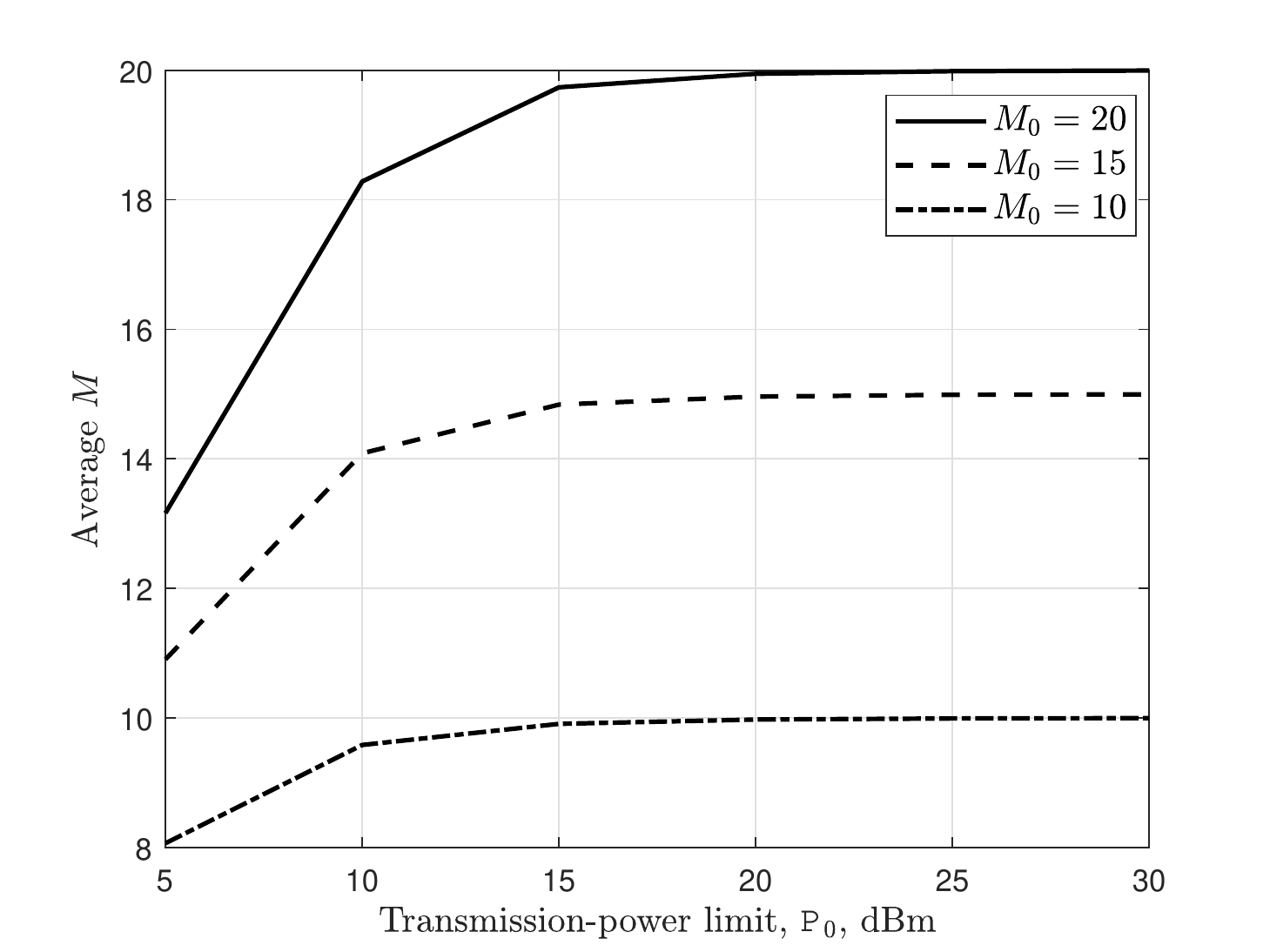}
	%	\vspace{-0.5cm}	
	\caption{The average number of plants under remote coding-free control versus the transmission-power limit, where $A=1.1$.}
	\label{fig:ave_num}
%	\vspace{-0.3cm}
\end{figure}

In Fig.~\ref{fig:ave_num}, we consider Rayleigh fading channel for each plant with average channel power gain $10^{-4}$, and plot the average number of plants chosen for coding-free control (i.e., average $M$) with different transmission-power limit $\mathsf{P}_0$ and number of plants $M_0$ by running $10^6$ channel realizations (control processes) and calculating $M$ using \eqref{c2}.
We see that the average number of plants in coding-free control is almost equal to the total number of plants $M_0$ when the transmission-power limit $\mathsf{P}_0\geq 20$~dBm.
%When the number of plants increases from $2$ to $6$, the transmission-power limit should be increased by $5$~dB to achieve the same control-outage probability.
%Recall that when the channel quality can lead to a control outage, in practice, the controller can switch to the mode of coding-based control as suggested in Fig.~\ref{fig:coding}.

\subsection{Optimal Coding-Free Control of Multi-Plant Systems in Fast-Fading Channels} \label{sec:num_fast}
\begin{figure}[t]
	\renewcommand{\captionfont}{\small} \renewcommand{\captionlabelfont}{\small}
	\centering
	%	\vspace{-0.3cm}	
	\includegraphics[scale=0.63]{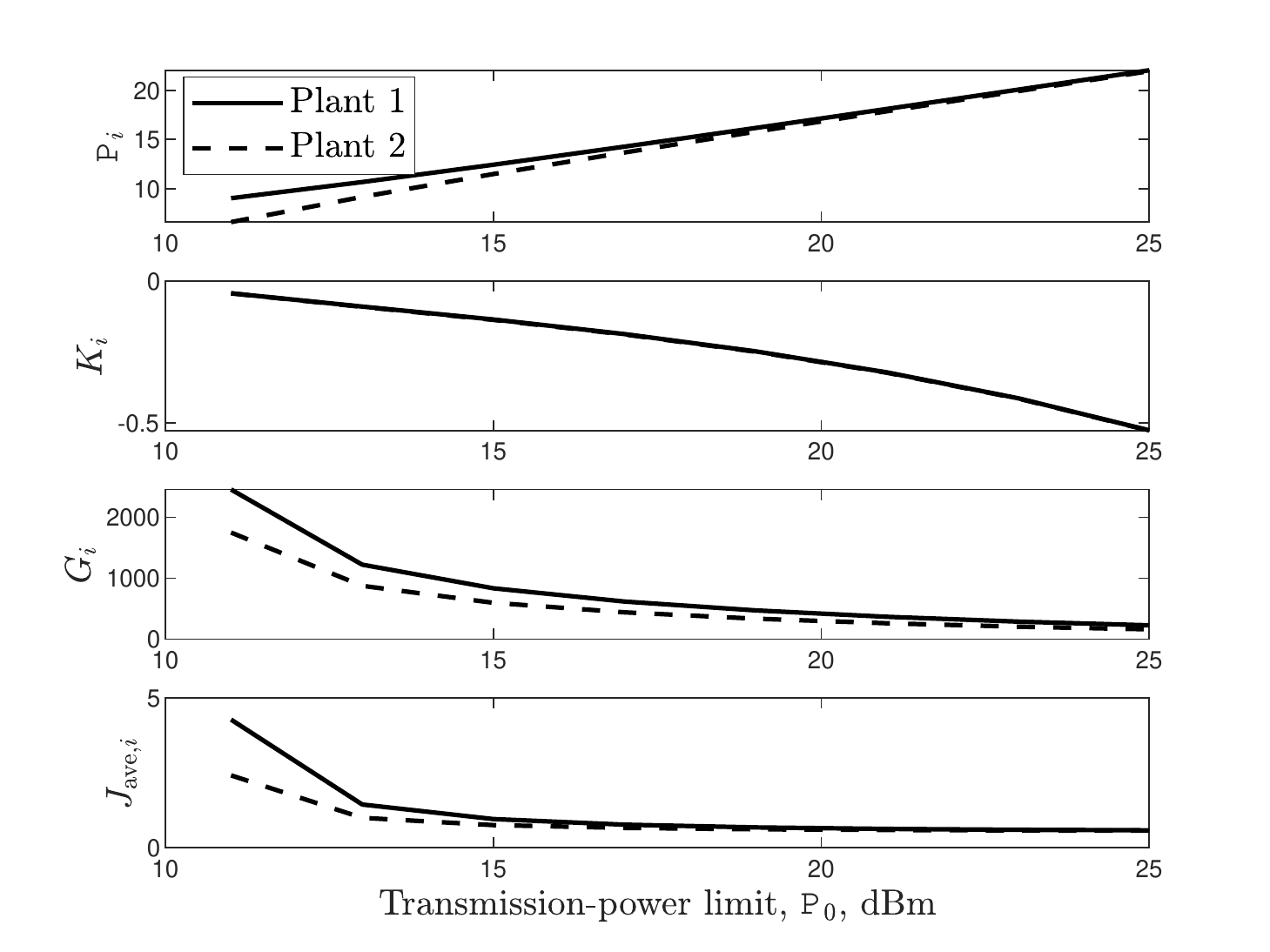}
	%	\vspace{-0.5cm}	
	\caption{The optimal control parameters and costs of the plants of the coding-free control method with different transmission-power limit in fast-fading channels.}
	\label{fig:optimal_fast}
%	\vspace{-0.3cm}
\end{figure}

In Fig.~\ref{fig:optimal_fast}, considering the fast-fading scenario, we plot the optimal allocated transmission power $\mathsf{P}_i$, control factor $K_i$, actuator factor $G_i$ and the average costs of a two-plant system of the optimal coding-free control policy, where the variances of the channel coefficients are $\sigma^2_{h,1}= 10^{-4}$ and $\sigma^2_{h,2}= 4 \times 10^{-4}$, respectively.
We see that it is not possible to stabilize both the plants when the transmission-power limit, $\mathsf{P}_0$, is less than $11$~dBm.
When $\mathsf{P}_0>11$~dBm, the optimal allocated power to each plant increases, while the optimal controller and actuator factors and the average cost of each plant decreases with the increasing of the transmission-power limit $\mathsf{P}_0$.
Comparing with the slow-fading case in Fig.~\ref{fig:optimal_parameter}, although the general trend of the curves are the same, the average costs of the plants in the fast-fading case are much higher than that in the slow-fading case due to the fluctuation and uncertainty of the channel coefficients. 
%Also, we see that although the plant with a worse channel condition, i.e., plant~1, is allocated with a larger transmission power and has a larger actuator factor than that of plant~2, the average cost of plant~1 is still larger than plant~2.
%Also, we see that the controller factor of plant 1 is larger and smaller than plant 2 in the low (e.g., $\mathsf{P}_0<8$~dBm) and high SNR regimes, respectively.

\section{Concluding Remarks}\label{sec:con}
{We have proposed a coding-free control method for  single-controller-multi-plant WNCS under both slow and fast fading channels. 
We have formulated and solved a joint controller-actuator design problem to optimize the sum cost functions of multiple plants, subject to the controller power limit and the plant stabilization conditions.
We have derived the optimal controller-actuator parameters in a closed form, though the original problem is non-convex. 
Furthermore, we have rigorously derived the condition on the existence of a coding-free control protocol that can stabilize the plants, in terms of the transmission-power limit and the channel conditions.
The insights brought by our results of coding-free WNCS in both the slow and fast fading scenarios are consistent with the intuition from conventional coding-free communication systems: the optimal power allocation policy is of a channel-inversion type, and the coding-free control method can provide better control-system performance than the conventional coding-based methods for a practical range of SNRs.

For future work, we will consider the design of coding-free control for more practical multi-state (vector) plant systems rather than scalar systems. This will introduce a new transmission-scheduling problem of the control signals of different plant states. For example, the state that is far from the preset value should have a higher priority to be controlled than the one that is close to the preset value.
We will also extend the coding-free control method into a more general multi-antenna setting, where both the controller and the actuators are equipped with multiple antennas. It will be interesting to investigate the optimal precoding design of the multi-antenna system for minimizing the cost function of the WNCS.
Moreover, the energy consumption of actuators for applying control actions can be added into the cost function of the WNCS, which thus captures both the control performance of the WNCS and the energy consumption for control. Based on the new cost function, we will design the optimal energy-efficient control policy of the WNCS.
Furthermore, one can think of an adaptive scheme based on the following tradeoff: the coding-free communication method has a low latency with noisy reception, while the coding-based one has longer latency but a greater accuracy. We will investigate a new transmission scheme that is able to adaptively switch between coding-free control mode and coding-based control mode based on the channel conditions and the system status. 

%
% to analyze the effect of tradeoff among latency and accuracy on the control performance of WNCSs.
%
% instead of focusing on coding-free control
%
%due that coding-free is low latency with noise reception, while coding-based is long latency but more accurate. So, can authors of this paper give an adaptive transmission scheme to analyze the trade-off among latency and accuracy
}

\balance

\ifCLASSOPTIONcaptionsoff
%\newpage
\fi
%
%\bibliographystyle{IEEEtran}
%\bibliography{IEEEabrv,cite}

\end{document}